\documentclass[%
 aip,
 amsmath,amssymb,
 reprint,%
]{revtex4-1}

\usepackage{graphicx}
\usepackage{dcolumn}
\usepackage{bm}

\usepackage[utf8]{inputenc}
\usepackage[T1]{fontenc}
\usepackage{mathptmx}
\usepackage{etoolbox}
\usepackage{xcolor}
\usepackage{titlesec}

\titleformat{\paragraph}[runin]{\normalfont\normalsize\bfseries}{\theparagraph}{1em}{}
\titlespacing*{\paragraph}{0pt}{3.25ex plus 1ex minus .2ex}{1em}

\makeatletter
\def\@email#1#2{%
 \endgroup
 \patchcmd{\titleblock@produce}
  {\frontmatter@RRAPformat}
  {\frontmatter@RRAPformat{\produce@RRAP{*#1\href{mailto:#2}{#2}}}\frontmatter@RRAPformat}
  {}{}
}%
\makeatother
\begin{document}

\preprint{AIP/123-QED}

\title{Effects of confinement, impinging shock deflection angle, and Mach number on the flow field of a supersonic open cavity}
\author{Sreejita Bhaduri}\affiliation{%
Department of Aerospace Engineering, Indian Institute of Technology Kanpur, 208016, Kanpur, India
}

\author{Mohammed Ibrahim Sugarno}
 \affiliation{%
Department of Aerospace Engineering, Indian Institute of Technology Kanpur, 208016, Kanpur, India
}
\author{Ashoke De}%
 \email{ashoke@iitk.ac.in.}
 \affiliation{%
Department of Aerospace Engineering, Indian Institute of Technology Kanpur, 208016, Kanpur, India
}
\affiliation{ 
Department of Sustainable Energy Engineering, Indian Institute of Technology Kanpur, 208016, Kanpur, India
}%


\begin{abstract}
Cavities exhibit inherent self-sustaining oscillations driven by the coupling between their hydrodynamic and acoustic properties. In practical applications, cavities are often placed within confinements that introduce compression waves, significantly influencing their primary flow characteristics. The oscillations in cavities have widespread applications, such as in fuel-air mixing, heat exchangers, and landing gears However, when resonance occurs, these oscillations can lead to structural failures. Therefore, understanding cavity oscillations under diverse geometrical configurations and flow conditions is essential.  The present study examines the impact of top wall confinement on an open cavity with a length-to-depth ratio (L/D) ratio of 3 at Mach 1.71, along with the effects of varying deflection angles on flow characteristics and the influence of an increased Mach number on configurations with the highest and lowest oscillation frequencies. A three-dimensional numerical investigation is carried out, employing large eddy simulations within the OpenFOAM framework. We analyze the flow fields through the spatial variation of density over time. Fast Fourier Transformation and Wavelet Transformation reveal the frequency content from unsteady pressure signals and illustrate its evolution over time under different conditions. Additionally, reduced-order modeling provides a better understanding of the relationship between frequencies and flow structures of the cavity. Results from these analyses demonstrate that top wall confinement increases oscillation frequency, while greater deflection angles introduce Kelvin-Helmholtz instability in the flow field, reducing the frequency. An increase in the Mach number to 2, further intensifies instability, substantially affecting oscillations. 
\end{abstract}

\maketitle

%
\section{\label{sec:level1} Introduction:}
Cavities have been the subject of extensive research for decades, driven by their broad applications and the complex physics they encompass. In the aerospace industry, cavities are utilized for various purposes, including weapon bays, fuel tanks, and resonator-type nozzles \cite{sahoo2005film,emmert2009numerical,johnson2010instability,chakravarthy2018analytical,devaraj2020experimental,saravanan2020isolator,sekar2020unsteady,krishnamurty1955acoustic}. Their intrinsic oscillations are crucial for mixing air with fuel in propulsion systems like scramjet combustors and are vital for thermal management in spacecraft and aircraft \cite{charwat1961investigation,emery1969recompression}. Cavities also play a significant role in enhancing heat transfer by increasing surface area in heat exchangers, improving cooling in electronics and cryogenic systems, and efficiently capturing solar radiation in solar energy applications \cite{swift2017thermoacoustics,choi1995enhancing,zalba2003review}. Additionally, they act as thermal barriers in building insulation and enable efficient coolant flow in nuclear reactors.

The flow characteristics of cavities largely determine their applications, with the length-to-depth ratio (L/D) playing a critical role in defining the flow type. Stallings \cite{stallings1987experimental} classified supersonic cavities with an L/D ratio of less than 10 as deep cavities, while those with an L/D ratio greater than 13 are categorized as shallow cavities. Deep cavities exhibit an open flow type, in which the shear layer flows over the cavity due to the high pressure ahead of the rear face venting into the low-pressure area downstream of the front face. Shallow cavities display a closed flow type in which the shear layer expands over the leading edge, impinges on the cavity floor, and exits ahead of the trailing edge, similar to the flows over forward and backward-facing steps.The cavity transits from open to closed flow when the L/D ratio ranges between 10 and 13. In this transitional flow type, the shear layer strikes the cavity floor downstream of the front edge, turns, and exits through the trailing edge, with the exit shock and impinging waves coalescing into a single wave. These flow classifications, particularly in supersonic and transonic regimes, depend on the Mach number and the cavity's depth-to-height ratio \cite{plentovich1993experimental,tracy1992measurements}.
The features of open and closed cavities change significantly from one another due to variations in their flow. There are no acoustic tones in closed cavities, according to Lawson and Barakos \cite{lawson2011review}.On the other hand, open cavities have a unique acoustic signature that is defined by the following characteristics:
\begin{itemize}
\item{Low-energy broadband noise produced by turbulent fluctuations, shear layer, and freestream.}
\item{Discrete tones, often referred to as Rossiter modes \cite{rossiter1964wind}, are generated by interactions such as shear-wall, shock shear layer, vortex-vortex, vortex-wall, or vortex-shear layer.}
\end{itemize}
Researchers like Heller \cite{heller1971flow,heller1975physical,heller1996cavity} and Krishnamurthy \cite{krishnamurty1955acoustic} briefly studied open supersonic cavities. Their studies indicate that in these cavities, the incoming shear layer separates from the leading edge and spans across the cavity. Periodic waves of expansion and compression at the trailing edge scavenge mass into and out of the cavity. A pressure wave is created close to the region in the cavity floor, where the mass flows as it enters the cavity through the aft wall. This pressure wave travels upstream, intensifies along the way, and perturbs the separating shear layer at the leading edge. This results in disturbances in the shear layer, which are convected downstream to impinge the trailing edge. A feedback loop is, thus, produced by this hydrodynamics and acoustics coupling.
\begin{figure*}

	\includegraphics[scale=0.45]{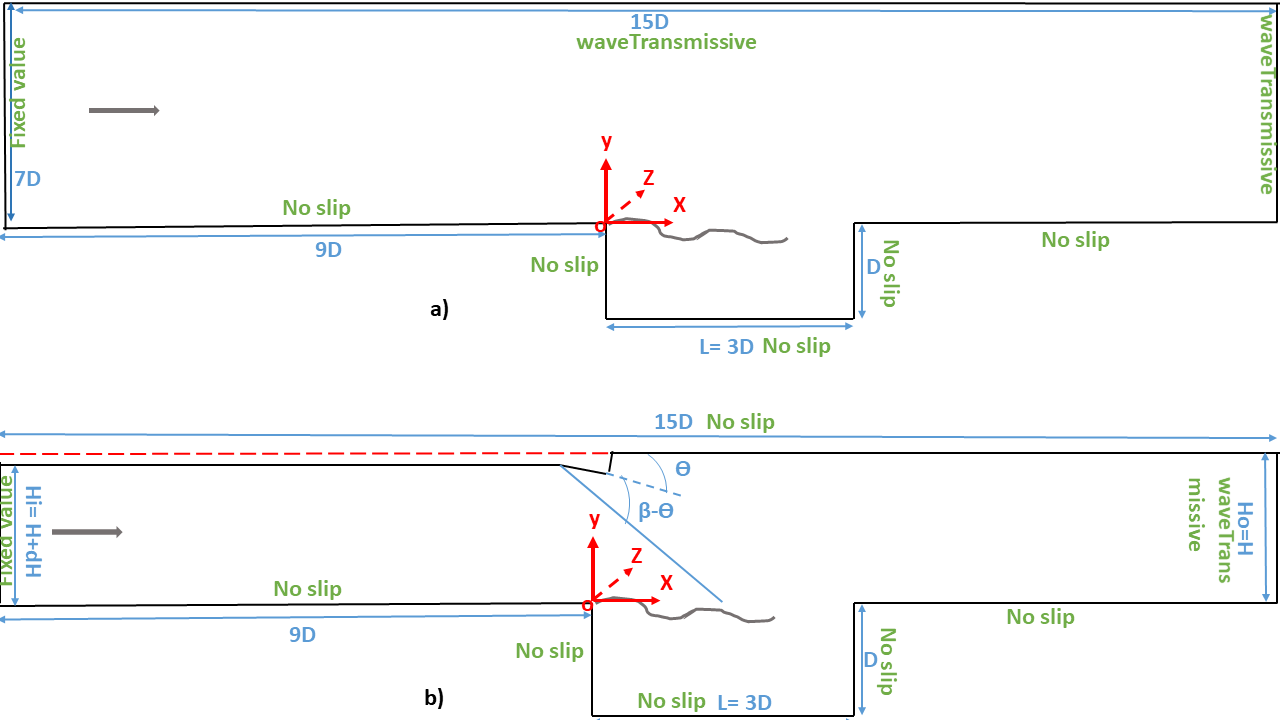}
    \centering

    \caption{\label{fig:1}Schematic Diagram of the cavity in the X-Y plane at the mid-section of the Z-axis a) without b) with top wall confinement along with the boundary conditions for the simulations. The origin is at the leading edge of the cavity.}.

    \end{figure*}
    \begin{figure}

	\includegraphics[scale=0.35]{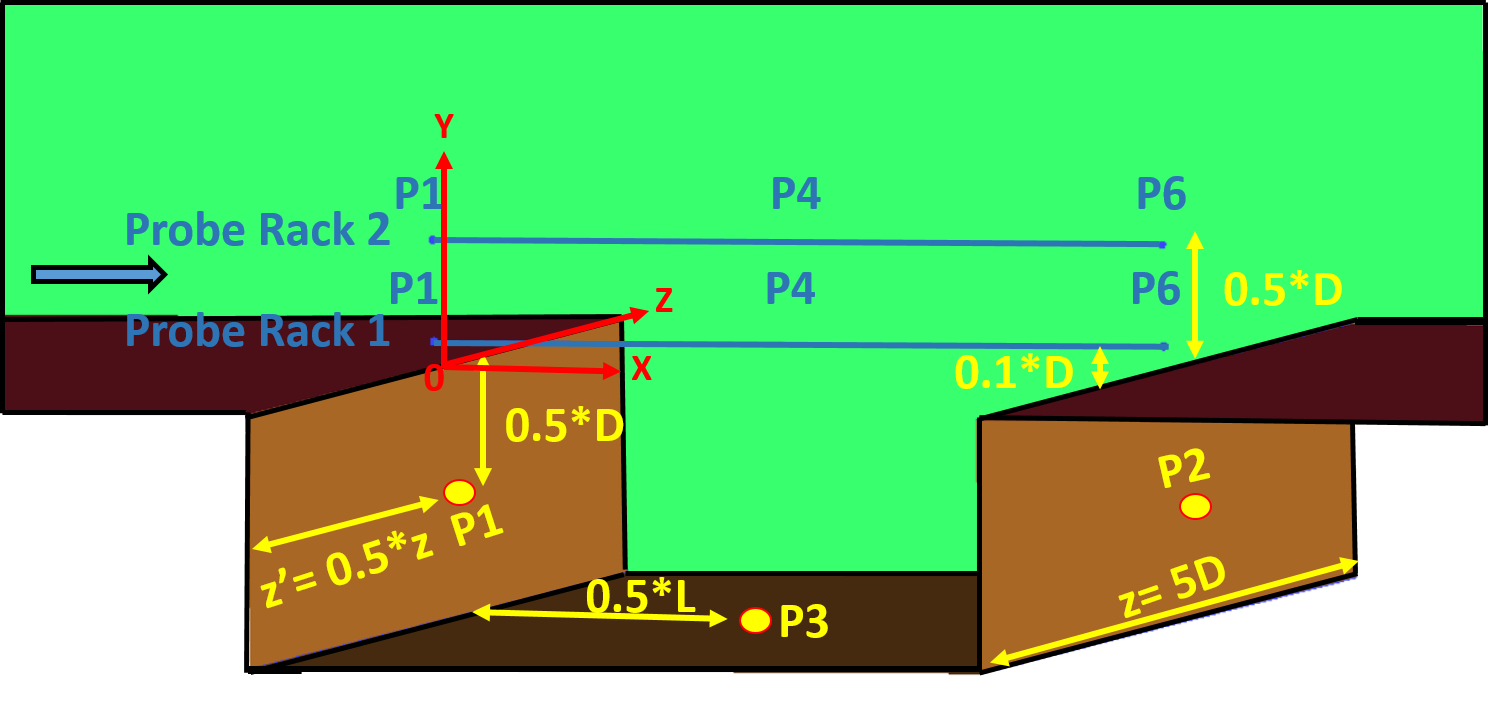}
    \centering

    \caption{\label{fig:2} Probe locations in the computational domain and walls of the cavity. These probes are placed in the mid-section of the z-axis.}

    \end{figure}

Supersonic flow over an open supersonic cavity is highly unsteady due to the inherent self-sustaining cavity oscillations \cite{rockwell1978self, woolley1974role}. Shear-layer instability, disturbance feedback, resonant wave conditions, and structural elasticity influence these oscillations either individually or in combination. These oscillations are categorized as fluid dynamic, fluid resonant, or fluid elastic oscillations, based on their origin. Fluid dynamic oscillations magnify unstable disturbances in the cavity shear layer, especially when the cavity length-to-acoustic wavelength ratio is minimal. Jet-edge oscillations caused by a free jet impingement on an edge resemble these oscillations. On the other hand, fluid-resonant cavity oscillations occur at frequencies where the acoustic wave wavelength is comparable to or lower than the cavity's characteristic length (L or W), and they strongly couple with resonant wave effects within the cavity. These oscillations may show transverse waves in deep cavities or longitudinal standing waves in shallow cavities, depending on the cavity's length-to-width ratio (L/W). Fluid-elastic oscillations occur when the cavity walls displace sufficiently to exert feedback control over shear-layer perturbations.

These self-sustaining oscillations in cavities can damage the structures in which they are stationed \cite{rockwell1978self,rowley2006dynamics}. Developing oscillation control techniques for their efficient usage in a variety of applications requires an understanding of cavity flow characteristics. Extensive research has focused on the primary flow characteristics of cavities and introduced both active and passive control mechanisms to reduce oscillations\cite{alam2007new,malhotra2016aft,lad2018experimental,jain2021aero,jain2023effects, panigrahi2019effects,lee2008passive}. These studies are mostly conducted in cavities without any top wall confinements. Confined cavities like those used in scramjet combustors \cite{sitaraman2021adaptive}, display numerous incidents and reflected shock waves, which are the result of multiple reflections between the confinement walls and the shear layer. The incident shock wave impinges the separating shear layer as it convects the disturbances downstream, significantly changing the flow field both within and above the cavity. Therefore, we must consider shock impingement on the shear layer to conduct a detailed study of cavity oscillations from the perspective of practical applications.

Karthick \cite{karthick2021shock} examined the effects of impingement of shock of varying strength on the shear layer by simulating unsteady flow past a confined supersonic cavity with an L/D ratio of 2. He conducted his study at a fixed Mach number of 1.71, utilizing Detached Eddy Simulations (DES) on a 2D geometry. The results highlighted the variation in wall pressure and oscillation frequency in the cavity due to changes in shock strengths interacting with the shear layer. Flow control techniques were developed, achieving a 58\% reduction in jet-column instability by using a confined supersonic wall-jet, where periodic large-scale structures significantly enhanced flow stability. The authors of the present article \cite{bhaduri2024flow} investigated passive control mechanisms for cavity oscillations in a 2D cavity with an L/D ratio of 3 at a Mach number of 1.71. The cavity was confined by a top wall with a deflection angle of $3.6^0$. Using Unsteady Reynolds-Averaged Navier-Stokes (URANS) simulations, they observed a notable suppression of cavity oscillations by placing sub-cavities at the front and aft walls. Oscillations are a defining feature of open cavities, crucial for their application across various fields. An accurate understanding of these dynamics necessitates three-dimensional simulations, which capture turbulence more effectively through mechanisms like vorticity stretching and complex interactions. The Mach number also significantly influences cavity flow characteristics, making it a critical parameter for comprehensive analysis. The aforementioned studies did not consider variations in Mach numbers or the three-dimensional effects in cavity flow fields. Therefore, this study investigates the influence of spanwise fluctuations on cavity flow by performing numerical simulations on an open cavity with an L/D ratio of 3, with specific focus on the following objectives: 
\begin{itemize}
\item {First, understanding the differences in the flow features and the frequency content of the cavity in the presence and absence of top wall confinement at a Mach number of 1.71.}
\item {Examining the effects of varying the shock's strength, impinging the separating shear layer, on the flow and oscillations of the cavity at Mach number 1.71.}
\item {Finally, assess how increasing the Mach number to 2 affects the flow characteristics of cavity configurations with the highest and lowest frequency content, as determined from the above investigations.}
\end{itemize}

We conduct these unsteady simulations on the compressible flow using Large Eddy Simulations within the OpenFOAM framework. Spectral analysis and numerical flow-field visualization provide in-depth insights into the configurations, while reduced-order modeling helps elucidate the dominant events in the flow fields.

The remainder of the article is organized as follows: Section \ref{sec:level2}
 covers the geometrical configuration, numerical methods, boundary conditions, grid independence study, and validation against experiments. Section \ref{sec:level3}
 presents the results and their physical explanation, detailing the variation in the flow field and frequency content with and without confinement, as well as impinging shock at Mach 1.71. This section further examines the effect of Mach number on the cavity dynamics with and without shock impingement. Section \ref{sec:conclusion} summarizes the key outcomes of this numerical investigation.

\section{\label{sec:level2} Computational Methodology:}
\subsection{Geometrical Configuration} \label{sec:geo}
Each cavity in this study measures 39 mm in length and 13 mm in depth (3D), with a spanwise extrusion of 5D to incorporate the effects of three-dimensionality in the flow. Figure \ref{fig:1}  shows the schematic diagrams of the X-Y plane at the Z-axis's mid-section (0.5z). The cavities are positioned 9D downstream from the intake to allow the flow to fully develop before entering the cavities. The domain is extended to 7D in the configuration with no wall confinement to ensure there is no reflection of the flow inside the domain across the top border (Figure\ref{fig:1}a) The top wall is positioned at a height ($H_i$) of 1.954D in the presence of confinement (Figure\ref{fig:1}b). The inlet and exit heights are equal ($H_i$=$H_o$), in the absence of shock-shear interaction. Ramps are positioned at the top wall with deflection angles ($\theta$) of $8.2^0$ and $13.78^0$.for shock impingement. The approaching supersonic flow is deflected by these ramps, resulting in the leading shock wave angle ($\beta$), $44.5^0$ and $52.87^0$respectively. For these two sets of deflection angles, the heights at the inlet are 1.969D and 2.021D, respectively, but the height at the outflow $H_o$ stays at 1.954D.  

 Figure \ref{fig:2} demonstrates the locations of the probes along the cavity walls and in the computational domain.  The probes are placed in the mid-section of the Z-axis (0.5z). P1, P2, and P3 are the designations for the three probes inside the cavity, at the middle of the front wall, aft wall, and base. We arranged six probes each in two racks at y=0.1D (probe rack 1) and 0.5D (probe rack 2) in the computing domain. Probe rack 2 is positioned slightly above the shear layer, whereas probe rack 1 is kept closer to it. These setups guarantee that important data is extracted from the flow field in every configuration that is being studied. P1 labels the first probes in each rack, while P6 labels the last. The locations of these probes are listed in Tables \ref{tab:table1} and\ref{tab:table2}. Every probe is positioned at z = 2.5 D, which is the mid-plane in the spanwise direction.

\begin{table}[ht]
    \centering
    \caption{\label{tab:table1}Location of the probes in probe rack 1 (y=0.1D).}
    \begin{ruledtabular}
    \begin{tabular}{cc}
        \textbf{Probe number (Pn)} & \textbf{Coordinates} \\
        \hline
        \mbox{n=1} & \mbox{(-D, 0.1D, 2.5D)} \\
        \mbox{n=2} & \mbox{(-0.5D, 0.1D, 2.5D)} \\
        \mbox{n=3} & \mbox{(0, 0.1D, 2.5D)} \\
        \mbox{n=4} & \mbox{(0.5D, 0.1D, 2.5D)} \\
        \mbox{n=5} & \mbox{(D, 0.1D, 2.5D)} \\
        \mbox{n=6} & \mbox{(2D, 0.1D, 2.5D)} \\
    \end{tabular}
    \end{ruledtabular}
\end{table}

\begin{table}[ht]
    \centering
    \caption{\label{tab:table2}Location of the probes in probe rack 2 (y=0.5D).}
    \begin{ruledtabular}
    \begin{tabular}{cc}
        \textbf{Probe number (Pn)} & \textbf{Coordinates} \\
        \hline
        \mbox{n=1} & \mbox{(-D, 0.5D, 2.5D)} \\
        \mbox{n=2} & \mbox{(-0.5D, 0.5D, 2.5D)} \\
        \mbox{n=3} & \mbox{(0, 0.5D, 2.5D)} \\
        \mbox{n=4} & \mbox{(0.5D, 0.5D, 2.5D)} \\
        \mbox{n=5} & \mbox{(D, 0.5D, 2.5D)} \\
        \mbox{n=6} & \mbox{(2D, 0.5D, 2.5D)} \\
    \end{tabular}
    \end{ruledtabular}
\end{table}

\begin{figure}

	\includegraphics[scale=0.4]{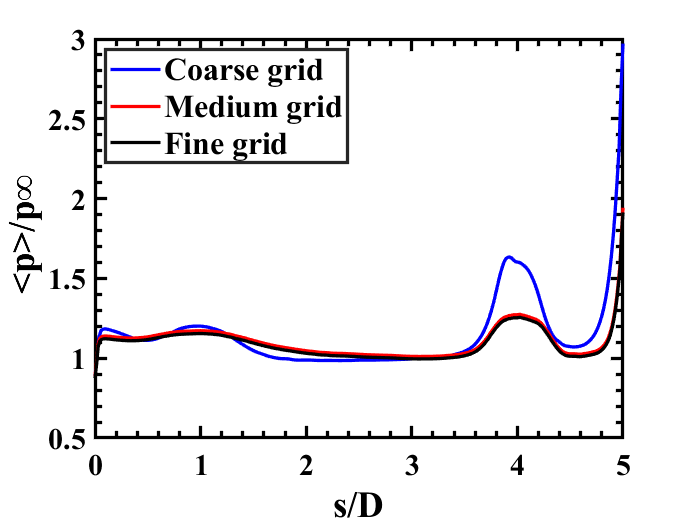}
   \centering

   \caption{\label{fig:3} Comparison of normalized cavity time-averaged wall pressure for all the grids used in the present simulation.}

   \end{figure}

    \begin{figure*}

	\includegraphics[scale=1.5]{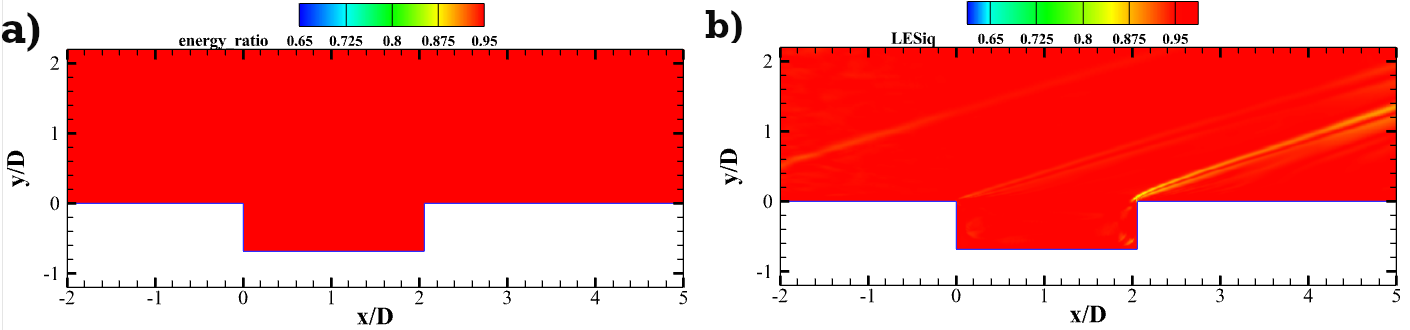}
    \centering

    \caption{\label{fig:4} Contours obtained for the medium grid of the present simulations for a) Turbulent Kinetic Energy to total energy ratio b) {LES\textsubscript{iq}}}.

    \end{figure*}

    \begin{figure}

	\includegraphics[scale=0.4]{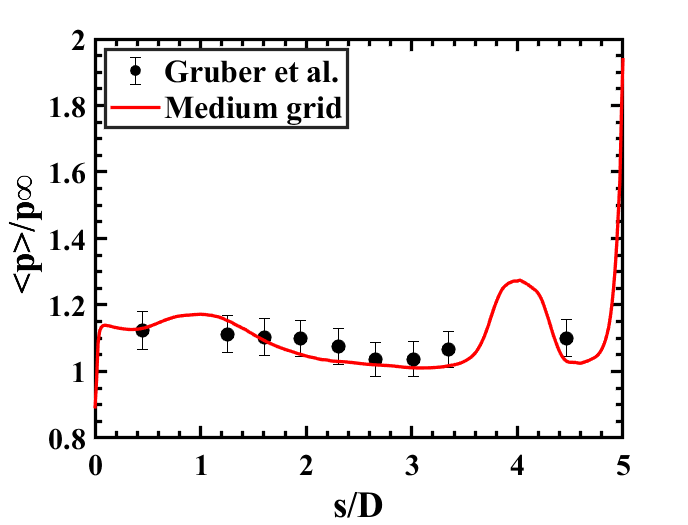}
    \centering

    \caption{\label{fig:5} Validation of the medium grid in the present simulation against the experimental data of Gruber et al.\cite{gruber2001fundamental}}.

    \end{figure}

\subsection{Numerical Methodology} \label{sub:nummerical}
The current study simulates flow via filtered Naviers Stokes (equations(\ref{eq:continuity},\ref{eq:momentum},\ref{eq:energy})) inherent to the Large Eddy Simulation (LES), which is frequently used to simulate turbulent flow fields. The filtering of Navier Stokes equations is based on Kolmogorov’s hypothesis. According to this theory, small-scale eddies are more ubiquitous, whereas large-scale ones are dependent on geometry. LES resolves large-scale eddies and uses a variety of sub-grid scale (SGS) models to model small-scale ones. The filtering mostly decomposes the velocity field into a resolved component and a subgrid-scale component \cite{moin1982numerical,piomelli1999large,cant2001sb,george2013lectures}. The filtered equations are standard Navier-Stokes equations with the SGS stress tensor. An eddy viscosity model further models the SGS stress tensor to close the equation. Averaged filtering equations are used in the majority of engineering applications to handle turbulent solutions. In the present study, we apply Favre-averaged (density-averaged) filtrated equations for the compressible flows. The working medium is assumed to abide by the ideal gas laws. The governing equations of the present simulations are listed below: 

\begin{equation}
\frac{\partial \overline{\rho}}{\partial t} +
\frac{\partial}{\partial x_i}\left[ \overline{\rho} \widetilde{u_i} \right] = 0
\label{eq:continuity}
\end{equation}

\begin{equation}
\frac{\partial}{\partial t}\left( \overline{\rho} \widetilde{u_i} \right) +
\frac{\partial}{\partial x_j}
\left[
\overline{\rho} \widetilde{u_j} \widetilde{u_i}
+ \overline{p} \delta_{ij}
- \widetilde{\tau_{ji}^{tot}}
\right]
= 0
\label{eq:momentum}
\end{equation}

\begin{equation}
\frac{\partial}{\partial t}\left( \overline{\rho} \widetilde{e_0} \right) +
\frac{\partial}{\partial x_j}
\left[
 \overline{\rho} \widetilde{u_j} \widetilde{e_0} +
 \widetilde{u_j} \overline{p} +
 \widetilde{q_j^{tot}} -
 \widetilde{u_i} \widetilde{\tau_{ij}^{tot}}
\right] = 0
\label{eq:energy}
\end{equation}

Where,
p, $\rho$,u, $e_0$,q are the pressure, density, velocity,total energy and heat flux of the system.$\tau$ represents the viscous stress. $(\overline)$ represents the time-averaged parameters whereas $(\widetilde{})$ represents the density averaged ones.To incorporate the effects of small-scale eddies,  viscous stress ($\tau$) and heat flux (q) are further split into laminar (lam) and turbulent (turb) components as: 
\begin{equation}
\widetilde{\tau_{ij}^{tot}} \equiv \widetilde{\tau_{ij}^{lam}} + \widetilde{\tau_{ij}^{turb}}
\label{eq:tau_tot}
\end{equation}

\begin{equation}
\widetilde{\tau_{ij}^{lam}} \equiv
\widetilde{\tau_{ij}} =
\mu
\left(
 \frac{\partial \widetilde{u_i} }{\partial x_j} +
 \frac{\partial \widetilde{u_j} }{\partial x_i} -
 \frac{2}{3} \frac{\partial \widetilde{u_k} }{\partial x_k} \delta_{ij}
\right)
\label{eq:tau_lam}
\end{equation}

\begin{equation}
\widetilde{\tau_{ij}^{turb}} \equiv
- \overline{\rho u''_i u''_j} \approx
\mu_t
\left(
 \frac{\partial \widetilde{u_i} }{\partial x_j} +
 \frac{\partial \widetilde{u_j} }{\partial x_i} -
 \frac{2}{3} \frac{\partial \widetilde{u_k} }{\partial x_k} \delta_{ij}
\right) -
\frac{2}{3} \overline{\rho} k \delta_{ij}
\label{eq:tau_turb}
\end{equation}

\begin{equation}
\widetilde{q_j^{tot}} \equiv \widetilde{q_j^{lam}} + \widetilde{q_j^{turb}}
\label{eq:q_tot}
\end{equation}

\begin{equation}
\widetilde{q_j^{lam}} \equiv
\widetilde{q_j} \approx
- C_p \frac{\mu}{Pr} \frac{\partial \widetilde{T}}{\partial x_j} =
- \frac{\gamma}{\gamma-1} \frac{\mu}{Pr} \frac{\partial}{\partial x_j}
  \left[ \frac{\overline{p}}{\overline{\rho}} \right]
\label{eq:q_lam}
\end{equation}

\begin{equation}
\widetilde{q_j^{turb}} \equiv
C_p \overline{\rho u''_j T} \approx
- C_p \frac{\mu_t}{Pr_t} \frac{\partial \widetilde{T}}{\partial x_j} =
- \frac{\gamma}{\gamma-1} \frac{\mu_t}{Pr_t} \frac{\partial}{\partial x_j}
  \left[ \frac{\overline{p}}{\overline{\rho}} \right]
\label{eq:q_turb}
\end{equation}

\begin{equation}
\overline{p} = \left( \gamma - 1 \right) \overline{\rho}
\left( \widetilde{e_0} - \frac{\widetilde{u_k} \widetilde{u_k}}{2} - k \right)
\label{eq:state}
\end{equation}

The eddy viscosity ($\mu_t$) is assessed in the current investigation using the Wall Adapting Local Eddy (WALE) viscosity model (equation:\ref{eq:mu}). The WALE model computes eddy viscosity based on the resolved velocity field's strain and rotation rates\cite{nicoud1999subgrid}. Compared to the Smagorinsky models, this predicts turbulent flows close to walls more accurately. The WALE solves the following equations to obtain $\mu_t$.

\begin{equation} \label{eq:mu}
\mu_{t} = \rho \Delta_s^2 \frac{(S_{ij}^{d} S_{ij}^{d})^{3/2}}{(\overline{S}_{ij} \overline{S}_{ij})^{5/2} + (S_{ij}^{d} S_{ij}^{d})^{5/4}}
\end{equation}

where,
\(\overline{S}_{ij}\) is the rate-of-strain tensor for the resolved scale defined by

\begin{equation}
\overline{S}_{ij} = \frac{1}{2} \left( \frac{\partial \overline{u_i}}{\partial x_j} + \frac{\partial \overline{u_j}}{\partial x_i} \right)
\end{equation}
 Filter width ($\Delta_s$) is defined by
\begin{equation}
\Delta_s = C_w V^{1/3}
\end{equation}
 \(C_w\) is a pure constant and in the present simulation is assumed to be 0.5.V is the cell volume.
 The velocity gradient tensor $\overline{g}_{ij}$ is
\begin{equation}
\overline{g}_{ij} = \frac{\partial \overline{u_i}}{\partial x_j}
\end{equation}
\begin{equation}
\overline{g}_{ij}^{2} = \overline{g}_{ik} \overline{g}_{kj}
\end{equation}
and the traceless symmetric
part of the square of the velocity gradient tensor($S_{ij}^{d}$):
\begin{equation}
S_{ij}^{d} = \frac{1}{2} \left( \overline{g}_{ij}^{2} + \overline{g}_{ji}^{2} \right) - \frac{1}{3} \delta_{ij} \overline{g}_{kk}^{2}
\end{equation}

A finite volume, density-based solver, "rhoCentralRK4Foam," \cite{greenshields2010implementation} solves the above governing equations in the OpenFoam \cite{jasak2007openfoam}framework. This solver handles high-speed flow difficulties such as strong shocks and contact discontinuities by discretizing convective fluxes using the central-upwind techniques of Kurganov and Tadmor\cite{kurganov2000new,weller1998tensorial, adityanarayan2023leading}. It employs an explicit, low-storage, third-order accurate, four-stage Runga Kutta approach for time integration and a second-order accurate central scheme for dissipative fluxes. We employ Sutherland's law to compute viscosity and the polynomial formula from the Joint Army-Navy-Air Force (JANAF) model to calculate the specific heat of air \cite{stull1965janaf,stull1974janaf}.

 The boundary conditions applied in the simulations are displayed in Figure 1. The flow initializes with a static pressure of 101325 Pa at a temperature of 189.29 K and velocities at  471 m/s (M = 1.71) and 551.576 m/s (M = 2). We introduce the initial fluctuations and turbulence using the "Klein" inflow generator \cite{klein2003digital}. At the outlet, all variables are extrapolated with the assumption of supersonic flow, and walls are assumed to be insulated with a "no-slip" condition.

 \subsection{ Validation}
The grid sensitivity and the flow solver are verified against the experimental data provided by Gruber et al \cite{gruber2001fundamental}. Their studies were conducted on a supersonic flow at a Mach number of 1.71 past an open cavity of L/D ratio of 3. The stagnation temperature and pressure were maintained at 300 k and 690 kPa, respectively, in their investigation.
\vspace{-2.5ex}
\subsubsection{Grid Independence}
Three computational grids are produced using the commercial program ICEM CFD \cite{ansys2011icem,ansys2016icem}  for optimization. For this purpose, the current simulations employ a coarse grid of 2.61 million, a medium grid of 5.172 million, and a fine grid of 10.223 million. Each of these grids is designed to keep the $y^+$ at 1 to resolve the boundary layer. The grid stretching in all three directions is devised with a suitable cell-size progression to maintain the aspect ratio below 30. Figure 3 shows that the results achieved with the medium and fine grids are almost the same. The literature states that if a grid resolves $80\%$  of the total kinetic energy, it is appropriate for LES \cite{arya2019effect}. The turbulent to total kinetic energy ratio in Figure 4 indicates the medium grid has a good resolution of the energy. We also evaluated the accuracy of the grid using LES\textsubscript{iq} \cite{celik2009assessment} and discovered that the medium grid is successful in capturing important aspects of turbulent flow. Hence, this grid is economical and efficient for validating solvers.

\vspace{-2.5ex}
\subsubsection{Solver Validation}
 The time-averaged pressure distribution along the cavity walls validates the results from the medium grid with Gruber's experimental data. Figure \ref{fig:5} shows the numerical results from the medium grid are within acceptable tolerance limits of $\pm 5\%$, when compared to the experimental data at all other locations, as suggested in the literature. The mismatch between the experimental and simulation results at x/D =4 is the result of the experiments' inadequate pressure taps which could not capture the rise in pressure at that location, as reported in the literature by Gruber.

\begin{figure*}

\includegraphics[scale=0.9]{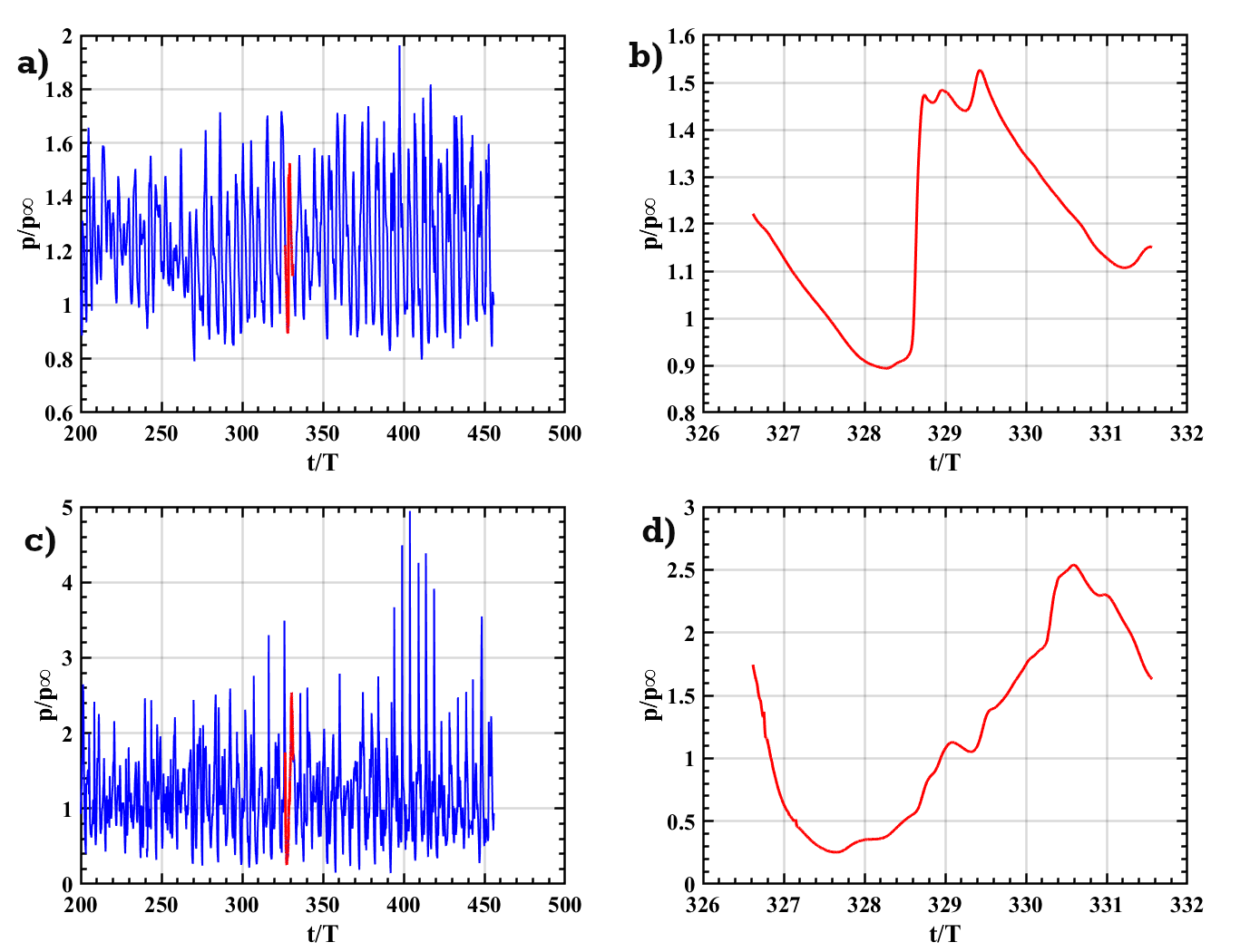}
    \centering

    \caption{\label{fig:6}  Temporal Variation of pressure normalized with the freestream pressure (p/$p_\infty$) at the a) front and b)aft walls of the cavity without confinement. The time is normalized with T (D/$U_\infty$=2.75e-5 s).}

    \end{figure*}
    \begin{figure*}

	\includegraphics[scale=0.31]{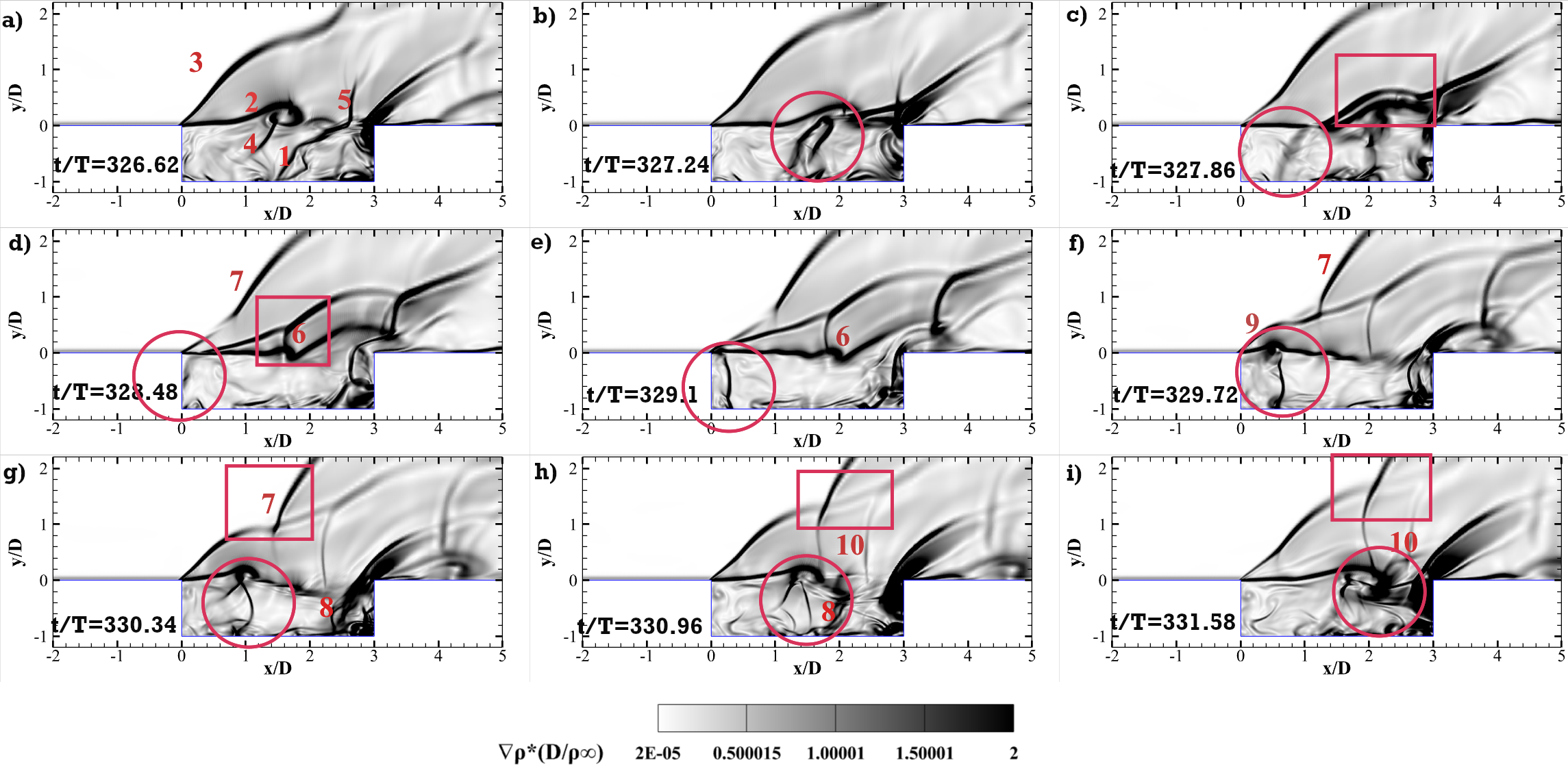}
    \centering
    \caption{\label{fig:7}  Normalised Density Gradient  ($\nabla\rho*(D/\rho_\infty$)) of the cavity without confinement for one complete cycle from the time step (t/T) of 326.62 (a) to 331.58(i) at an interval of 0.62. The numbered regions and their associated flow features are as follows: (1) the upstream traveling wave, (2) the perturbed separating shear layer from the previous cycle, (3) the separation shock, (4) the downstream traveling wave from the prior cycle, (5) the external wavefront accompanying the upstream wave, (6) disturbances in the perturbed shear layer, (7) compression waves at the leading edge, (8) the pressure wave of the next cycle, (9) the separation shock at the leading edge of the current cycle, and (10) disturbances generated in the present cycle convecting downstream. The red circle marks the flow features related to the feedback loop within the cavity. The red square demonstrates the flow features outside the cavity.  }

    \end{figure*}
     \begin{figure*}

	\includegraphics[scale=0.9]{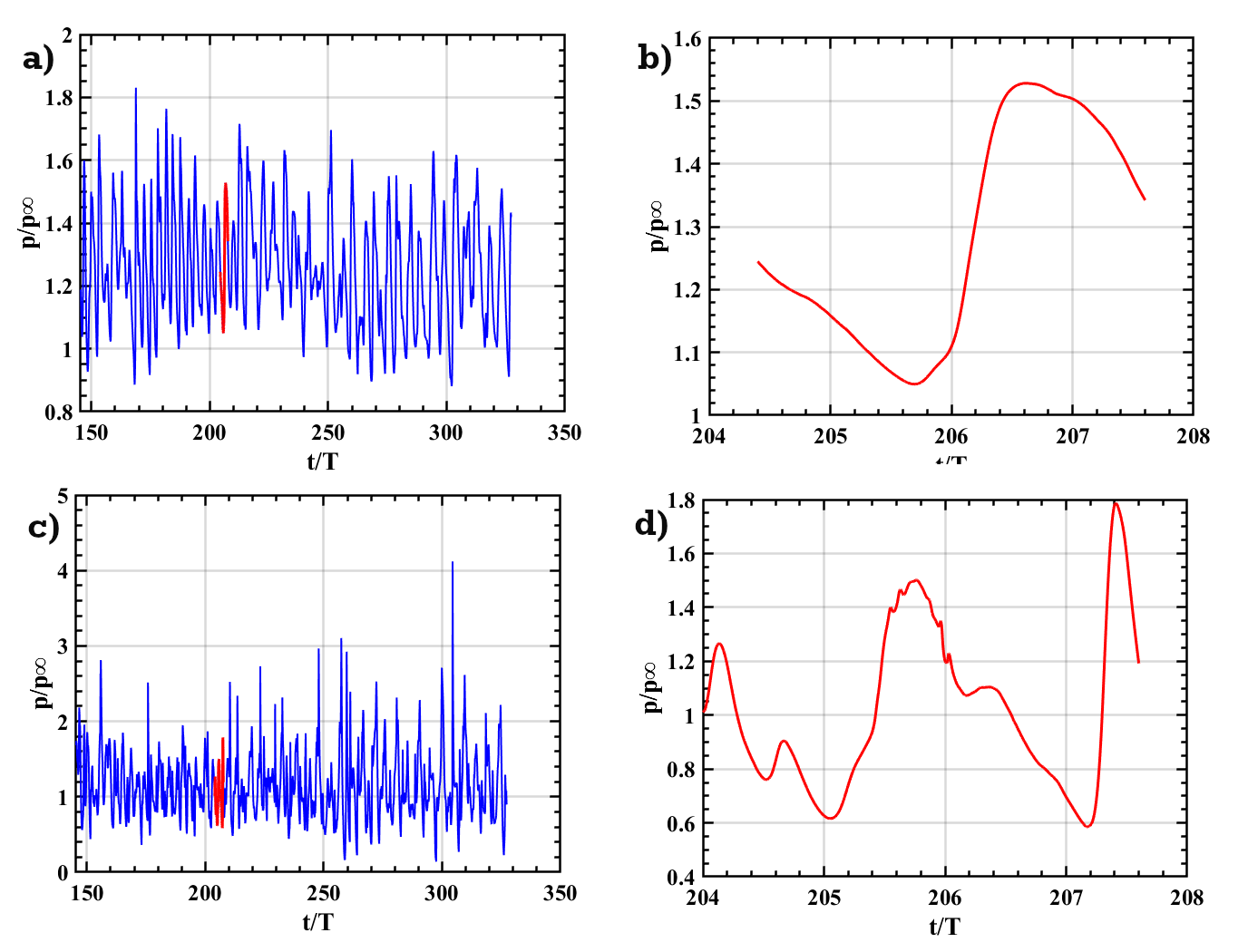}
    \centering

    \caption{\label{fig:8} Temporal Variation of pressure normalized with the freestream pressure (p/$p_\infty$) at the a) front and b)aft walls of the cavity with confinement deflected at $0^0$ The time is normalized with T (D/$U_\infty$=2.75e-5 s).}.
   \end{figure*}
 \begin{figure*}

	\includegraphics[scale=0.33]{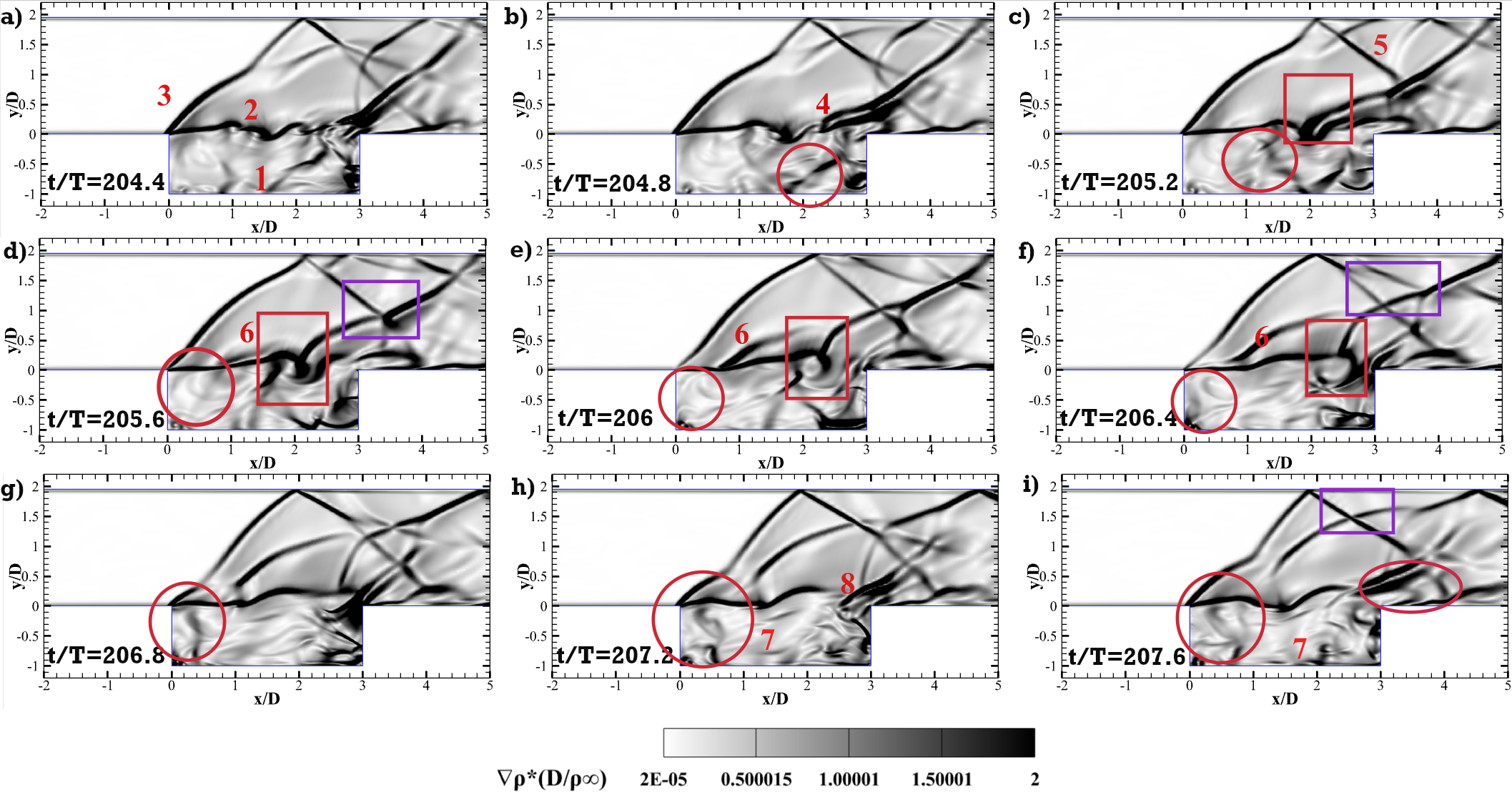}
    \centering

    \caption{\label{fig:9}Normalised Density Gradient  ($\nabla\rho*(D/\rho_\infty$)) of the cavity with the top wall deflected at $0^0$  for one complete cycle from the time step (t/T) 204.4 (a) to 207.6(i) at an interval   of 0.4. The numbered regions in the figure correspond to distinct flow features:  (1) the upstream traveling pressure wave from the current cycle, (2) the perturbed separating shear layer influenced by the previous cycle, (3) the separation shock, (4) the external wavefront trailing the upstream wave, (5) the reflected shock from the top wall, (6) the shear layer perturbations in the current cycle, (7) the upstream traveling waves of the next cycle, and (8) the interactions between downstream convecting disturbances and the compression waves. The red circles show the features inside the cavity. The red and blue square illustrates the multiple events outside the cavity.}

    \end{figure*}

    \begin{figure*}

	\includegraphics[scale=0.9]{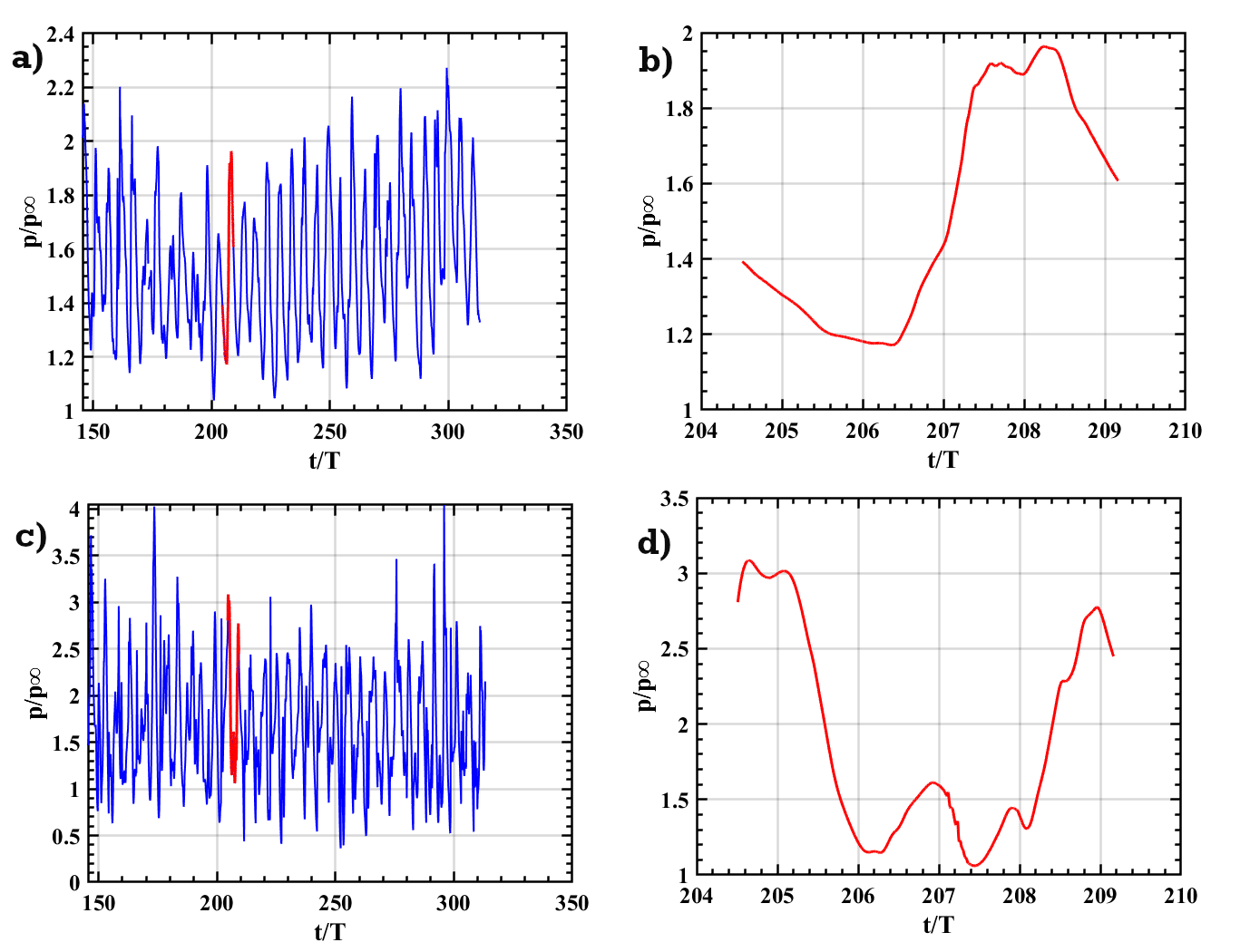}
    \centering

    \caption{\label{fig:10}  Temporal Variation of pressure normalized with the freestream pressure (p/$p_\infty$) at the a) front and b)aft walls of the cavity with the top wall of the deflection angle of $8.2^0$. The time is normalized with T (D/$U_\infty$=2.75e-5 s).}.

    \end{figure*}
     \begin{figure*}

	\includegraphics[scale=0.36]{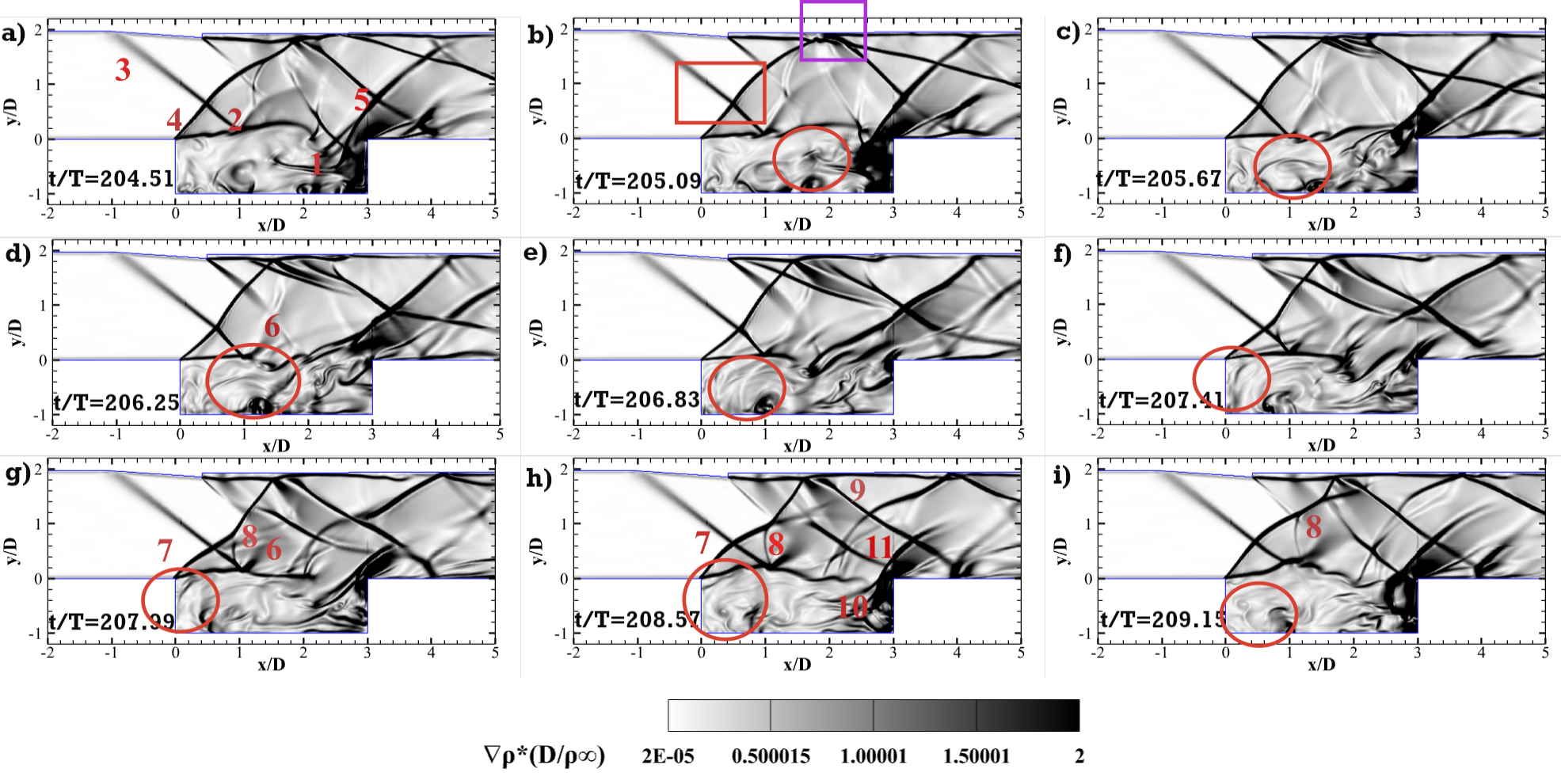}
    \centering

    \caption{\label{fig:11} Normalised Density Gradient  ($\nabla\rho*(D/\rho_\infty$)) of the cavity with the top wall deflected at $8.2^0$  for one complete cycle from the time step (t/T) 204.51 (a) to 209.15 (i) at an interval   of 0.58. Significant flow features are numbered as (1) the upstream traveling pressure waves,(2) the shear layer perturbed from the previous cycle, (3) the impinging shock wave generated at the deflection, (4) the separation shock at the leading edge, (5) trailing external wavefront, (6) disturbances created in the shear layer by the shock impingement, (7) the separation shock of the present cycle, (8)disturbances created at the impingement point by the perturbations convecting downstream, (9) shock reflected from the top wall, (10) \& (11) the trailing wavefront and the upstream traveling waves from the next cycle.  The red circles show the features inside the cavity. The red and purple squares illustrate the multiple events outside the cavity.}

    \end{figure*}
       \begin{figure*}

	\includegraphics[scale=0.9]{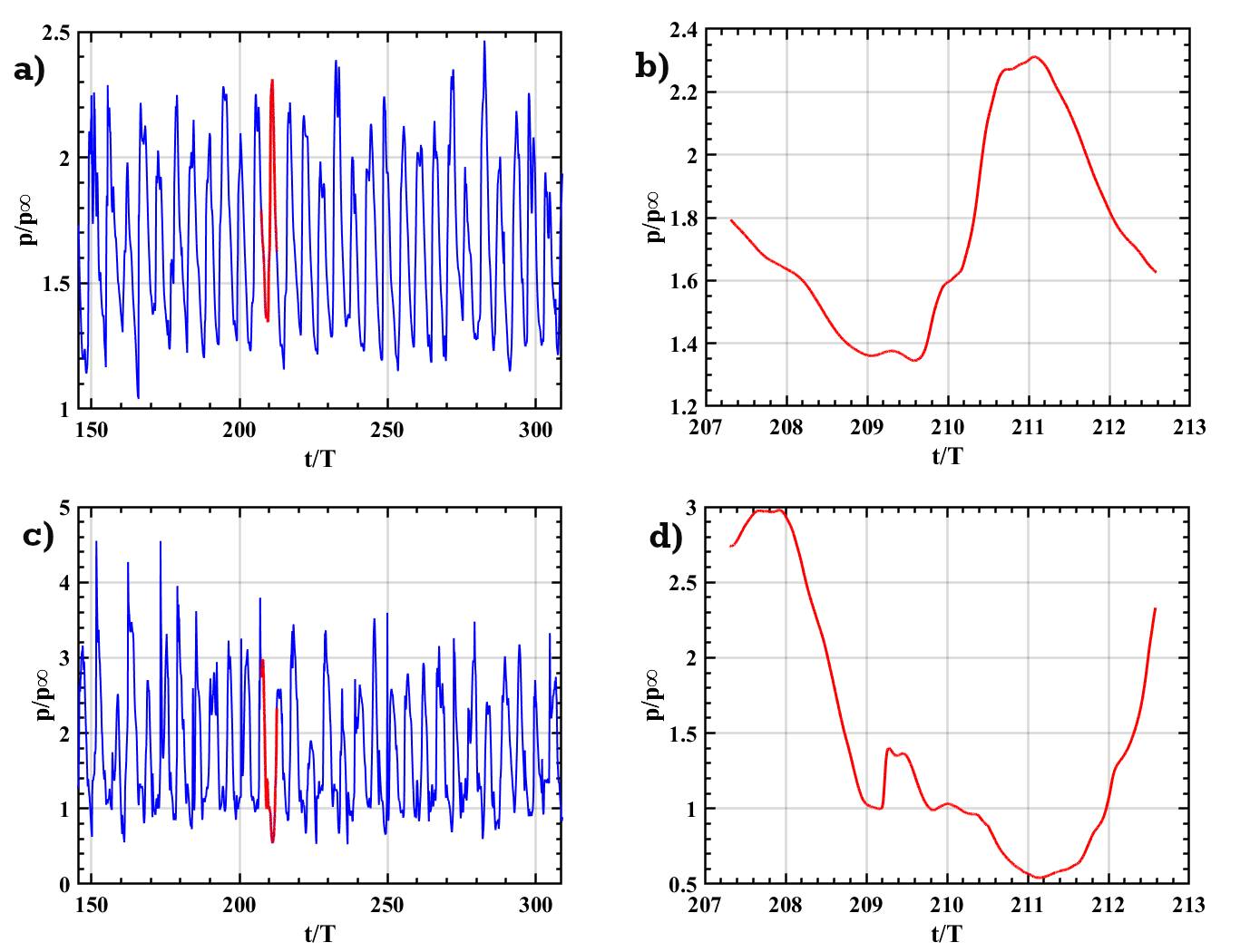} 
    \centering

    \caption{\label{fig:12}  Temporal Variation of pressure normalized with the freestream pressure (p/$p_\infty$) at the a) front and b)aft walls of the cavity with the top wall of the deflection angle of $13.78^0$. The time is normalized with T (D/$U_\infty$=2.75e-5 s).}.

    \end{figure*}
         \begin{figure*}

	\includegraphics[scale=0.36]{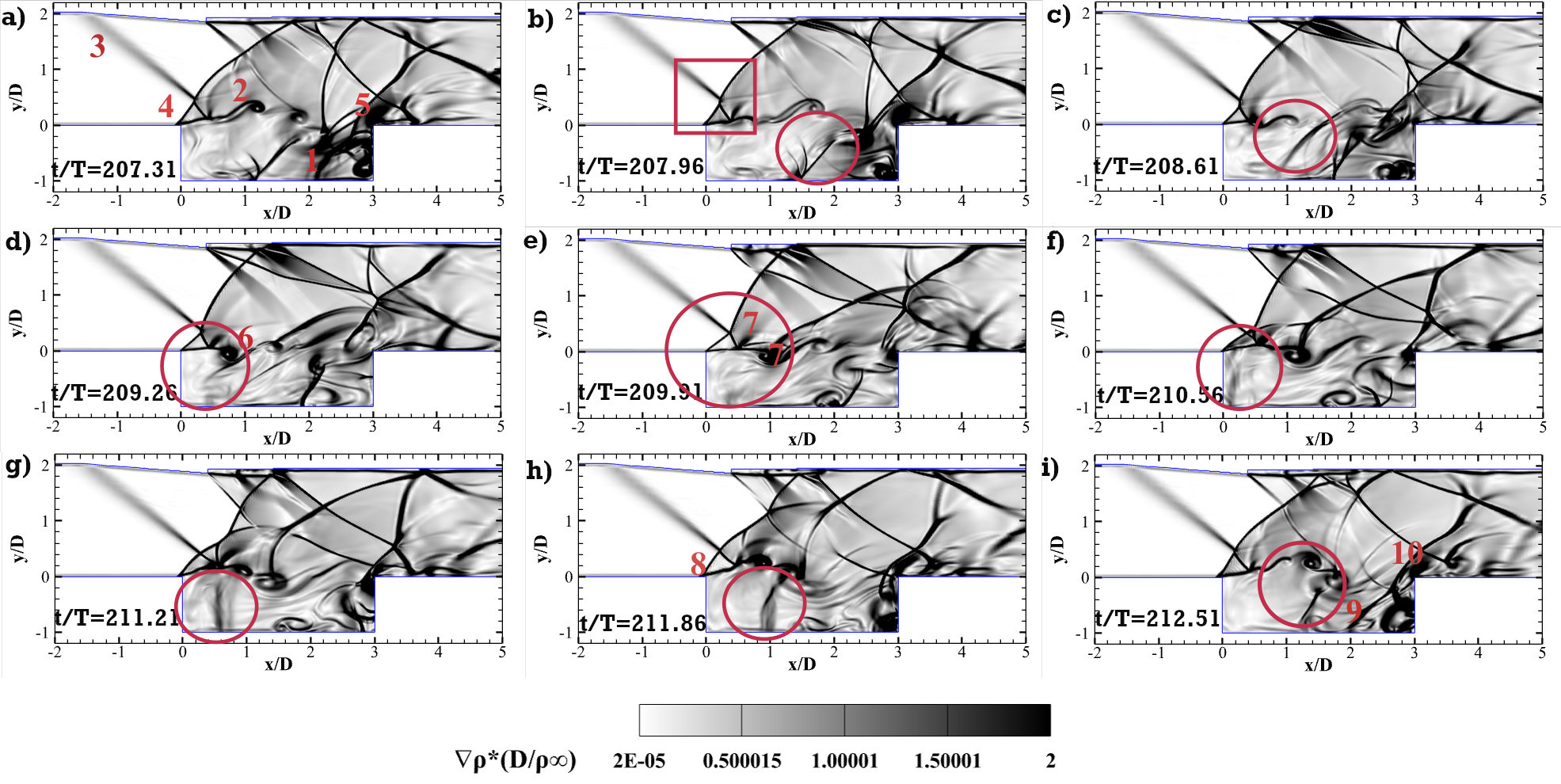}
    \centering

    \caption{\label{fig:13} Normalised Density Gradient  ($\nabla\rho*(D/\rho_\infty$)) of the cavity with the top wall deflected at $8.2^0$  for one complete cycle from the time step (t/T) 207.31 (a) to 212.51 (i) at an interval of 0.65. The salient features in the flow field are marked as: (1) the upstream traveling pressure waves (2)the K-H instability rolls in the perturbed shear layer, (3) the impinging shock wave from the deflected top wall, (4) the separation shock at the leading edge, (5) trailing wavefront, (6) downstream convecting disturbances from the previous cycle, (7) disturbances generated at the point of impingement, (8) separation shock of the next cycle, (9) \& (10) the upstream traveling wave and the external wavefront trailing it for the next cycle.  The red circles show the features inside the cavity. The red square illustrates the multiple events outside the cavity.}

    \end{figure*}

\section{\label{sec:level3} Results and Discussions:}
This section is divided into two subsections. In the first subsection, we examine how confinement and shock impingement on the shear layer influence the cavity's frequency content and flow field at Mach 1.71. The effect of increasing the Mach number to 2 on the flow field of the confined cavity configurations with the highest and lowest frequency content is discussed in the second subsection. We majorly utilise three methods in our study to comprehend the flow features emerging in the flow field: flow visualization by the spatial variation of density, spectral analysis to examine the frequency content of the system and its evolution over time; and dynamic orthogonal decomposition to understand the relate the flow structures to the various frequencies obtained in all the configurations.
\subsection{\label{effect1}Effect of confinement and shock-shear layer interaction on the cavity characteristics.}
\subsubsection{\label{sec:flow} Flow Field Visualisation}
 We examine the spatio-temporal variation of density for each case to analyze the change in the flow dynamics of the cavities of the various configurations. One inherent flow characteristic of all these arrangements is the periodic feedback loop that exists inside the cavity. Hence, we first identify one complete cycle in the time analysis of the normalized pressure (p/ $p_\infty$) obtained from the probes positioned at the front and aft walls of the cavities. Following this, we extract the density gradient contours for those specific time steps to elucidate the associated flow features. The density gradient magnitude ($\nabla\rho$) is normalized with the ratio of the depth of the cavity(D) to the freestream density ($\rho_\infty$). The time (t) is normalized with T, which is the ratio of the depth of the cavity (D) with the freestream velocity ($U_\infty$).

Figure \ref{fig:6}illustrates the variation of the normalized pressure at the front and aft walls of the cavity without any confinement. One complete cycle is identified between the normalized time (t/T) of 326.62 and 331.58. The pressure at the front wall is lower than the pressure at the aft wall, as can be seen from the pressure plots in Figure \ref{fig:6}. As a result, there is an influx of mass through the trailing edge as air flows from the aft wall to the front wall. The pressure at the front wall reaches its minimum, then increases sharply, as the flow advances from the aft to the front wall. It gradually decreases to the previous value initiating the next cycle. The pressure at the aft wall rises gradually over this course. This indicates that mass is being expelled from the trailing edge. The alternating inflow and outflow of mass through the trailing edge is a fundamental property of the cavity. The pressure at the aft wall again decreases towards the end of the cycle, facilitating the inflow of mass for the next cycle. 

Figure \ref{fig:7} demonstrates the flow field in the cavity during this time period. Figure \ref{fig:7}a shows the separating shear layer (2) is moving towards the trailing edge with a separating shock (3) emanating from the leading edge of the cavity and the upstream traveling wave(1) along with an external wavefront (5), moving towards the front wall. In this instant, the pressure wave (4) from the previous cycle moves downstream along with the disturbances in the excited shear layer(2). Their point of interaction has a definite vortex, which is encountered by (1) while traveling upstream (Figure \ref{fig:7}b). After this encounter, the upstream and downstream traveling waves move along their respective directions. This interaction, however, results in the generation of a disturbance (6)(figures \ref{fig:7}d-h), which travels downstream and dissipates in due course. In the subsequent time steps, the upstream traveling pressure waves (marked in a red circle), travel towards the leading edge and impinge on it. This wave while reflecting back, excites the separating shear layer (Figures \ref{fig:7} c-d). A vortex forms from the point of interaction of the shear layer with the reflected wave which convects downstream(Figure \ref{fig:7}e-i). A shock (9) is generated at the corner of the leading edge due to the shear layer separation. It interacts with the shock of the previous cycle (7) as the shear layer is traveling downstream to impinge on the aft wall (Figures\ref{fig:7}e-g). The reflected wave from this shock-shock interaction, impinges on the vortex in the shear layer (Figure\ref{fig:7}h), as it convects the perturbations toward the aft wall. This results in the disturbance (10), which also interacts with the pressure wave of the next cycle (8) which has developed in due course and is traveling upstream (Figures\ref{fig:7}f-h). As this shear layer impinges the aft wall, there is some outflow of mass, shown by the growing boundary layer downstream to the trailing edge (Figure\ref{fig:7}i). So, the flow visualization captures the feedback loop vividly.

The pressure variation at the cavity walls with top wall confinement at a deflection angle of $0^0$ is shown in Figure \ref{fig:8}. As observed in the preceding scenario, the pressure at the front wall is lower than the pressure at the aft wall at the start of the cycle.  This initiates the air movement from the rear wall to the front wall. Consequently, the mass enters through the trailing edge. The aft wall pressure, however, exhibits several fluctuations in a single cycle, in contrast to the pressure at the walls of the cavity without confinement. This suggests that there are multiple dynamics present at the aft wall in addition to the alternating input and outflow of the mass within the cavity. The pressure at the front wall and the rear wall approach the same value as they did at the start of the cycle. This marks the beginning of the subsequent cycle.
    
The upstream moving wave (1), the separating shear layer (2), and the separation shock (3) are shown at the leading edge in Figure\ref{fig:9}a. After interacting with the structures in the perturbed shear layer (Figures  \ref{fig:9}b-c), the trailing wavefront (4) of (1) bifurcates. This wave splits into two parts, one (6) trailing with (1) and the other traveling downstream with the shear layer (Figures \ref{fig:9}d-f). Before hitting the boundary layer downstream of the trailing edge, the separation shock (3) travels to the top wall and reflects as a shock wave (5). This shock wave then interacts with the pressure wave (1)'s external wavefront (4). The point of interaction fluctuates, as depicted in Figures \ref{fig:9}b-h, as the external wavefront moves downstream after splitting from the upstream waves before the latter makes contact with the leading edge. As the reflected shock from the top impinges the boundary layer at the trailing edge with varying strength depending on its interaction with the disturbances from the excited shear layer, near the aft wall, the pressure varies. The motion of the wave moving upstream within the cavity is also depicted in these figures. The separating shear layer bulges out as it interacts with the reflected wave (Figures  \ref{fig:9}g-i). Along with the disturbances originating from the point of interaction between the reflected wave and the separating shear layer (shown in red circles in Figures  \ref{fig:9}g–i), a separation shock also arises at the leading edge. As the pressure wave moves downstream inside the cavity, the trailing wavefront (6) separates from it and engages with the reflected shock from the upper wall. While the perturbations in the shear layer travel downstream to the trailing edge, the upstream traveling wave (7) of the next cycle is generated.
  
   
The confined cavity completes a cycle faster than the unconfined cavity, according to the flow visualization data for the cavity designs with and without top wall confinement. In comparison to the unconfined configuration, the confined cavity's aft wall experiences lower overall pressure. In addition, the pressure drop that occurs in the confined cavity at the end of each cycle, indicating the start of the subsequent cycle, is more pronounced and rapid than in the unconfined cavity. The faster pressure drop allows for a larger mass influx, which speeds up the feedback process. 
    \begin{figure*}

	\includegraphics[scale=0.75]{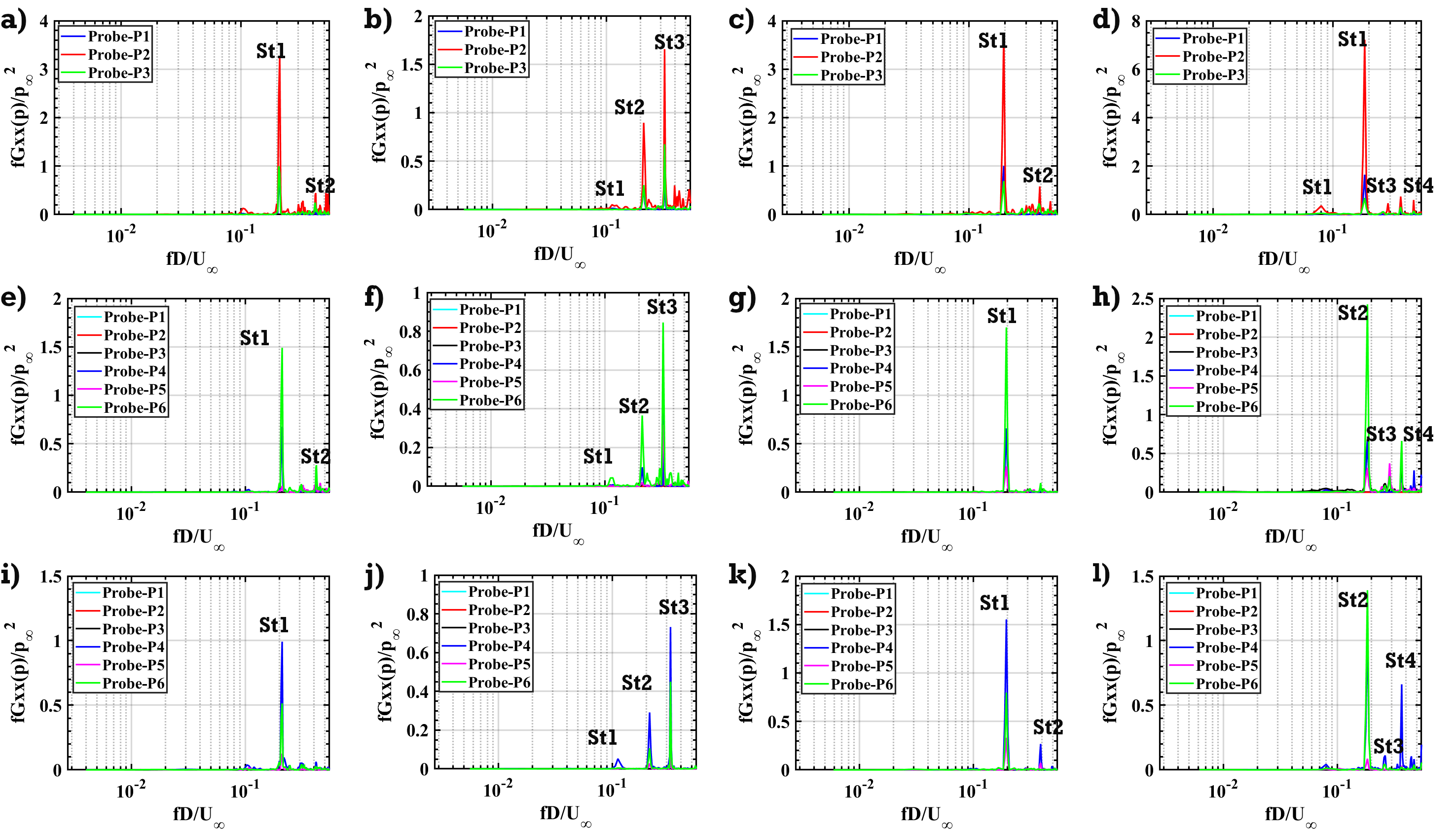}
    \centering
\caption{\label{fig:14} Normalised Power Spectral Density (PSD) (fGxx(p)/$(p_\infty)^2$) of the probes at the walls of the cavity, with the corresponding the Strouhal number (St)(fD/$U_\infty$) : (a) without confinement, and with confinement and  (b) no shock impingement, (c) with a flow deflection angle of $8.2^\circ$, and (d) with a flow deflection angle of $13.78^\circ$. For probe rack 1: (e) without confinement, and with confinement and (f) shock impingement, (g) with a flow deflection angle of $8.2^\circ$, and (h) with a flow deflection angle of $13.78^\circ$. For probe rack 2: (i) without confinement, and with confinement and (j) no shock impingement, (k) with a flow deflection angle of $8.2^\circ$, and (l) with a flow deflection angle of $13.78^\circ$.}

\end{figure*}
 \begin{figure*}
    \centering
    \includegraphics[scale=0.75]{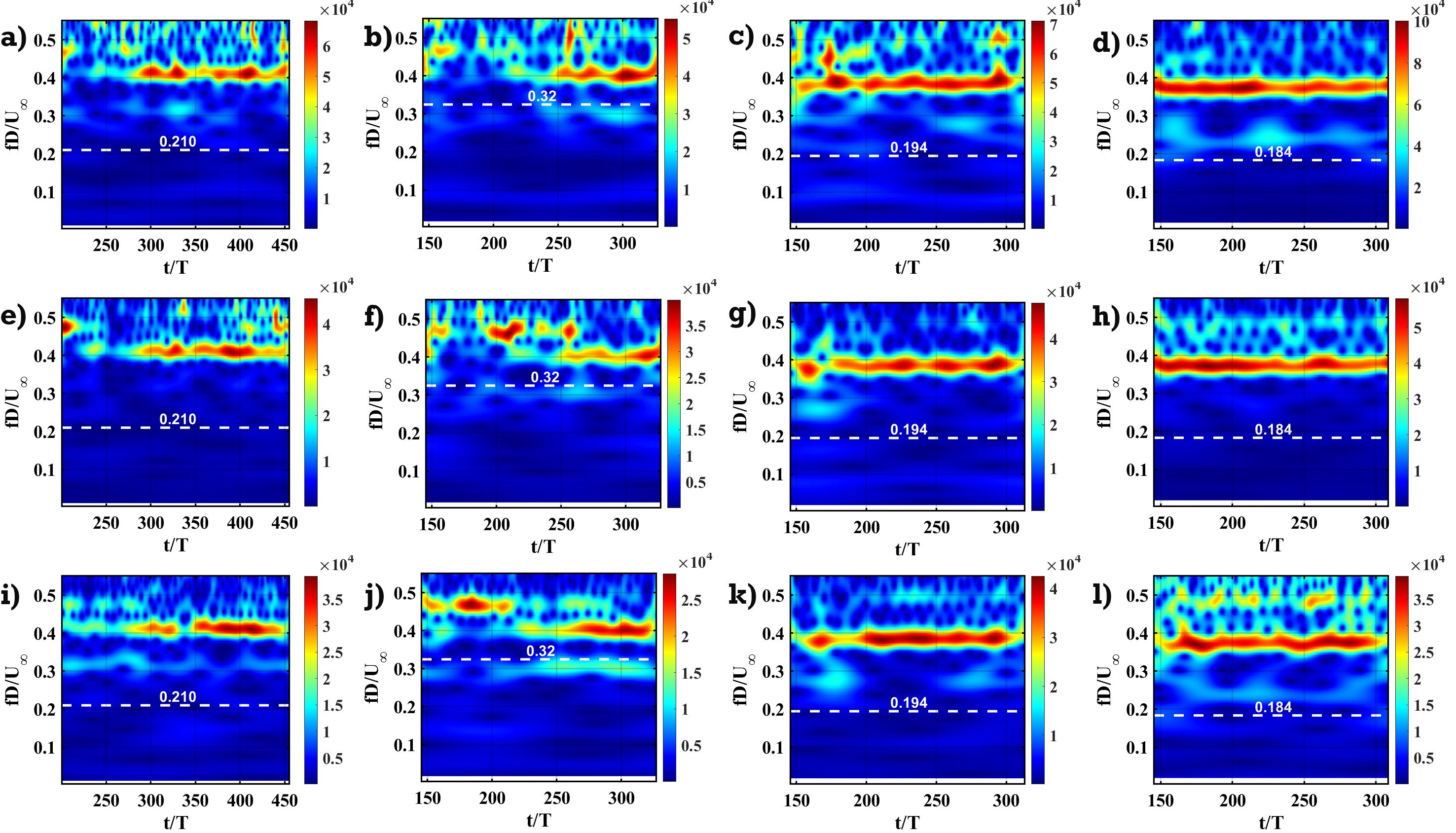} 
   \caption{\label{fig:15} Continuous Wavelet Transformation (CWT) of probes at the walls of the cavity, with the respective dominant Strouhal number across the time (t/T), marked: (a) without confinement, and with confinement and  (b) no shock impingement, (c) with a flow deflection angle of $8.2^\circ$, and (d) with a flow deflection angle of $13.78^\circ$. For probe rack 1: (e) without confinement, and with confinement and (f) shock impingement, (g) with a flow deflection angle of $8.2^\circ$, and (h) with a flow deflection angle of $13.78^\circ$. For probe rack 2: (i) without confinement, with confinement and (j) no shock impingement, (k) with a flow deflection angle of $8.2^\circ$, and (l) with a flow deflection angle of $13.78^\circ$.}

\end{figure*}
\begin{figure*}
    \centering
    \includegraphics[scale=0.75]{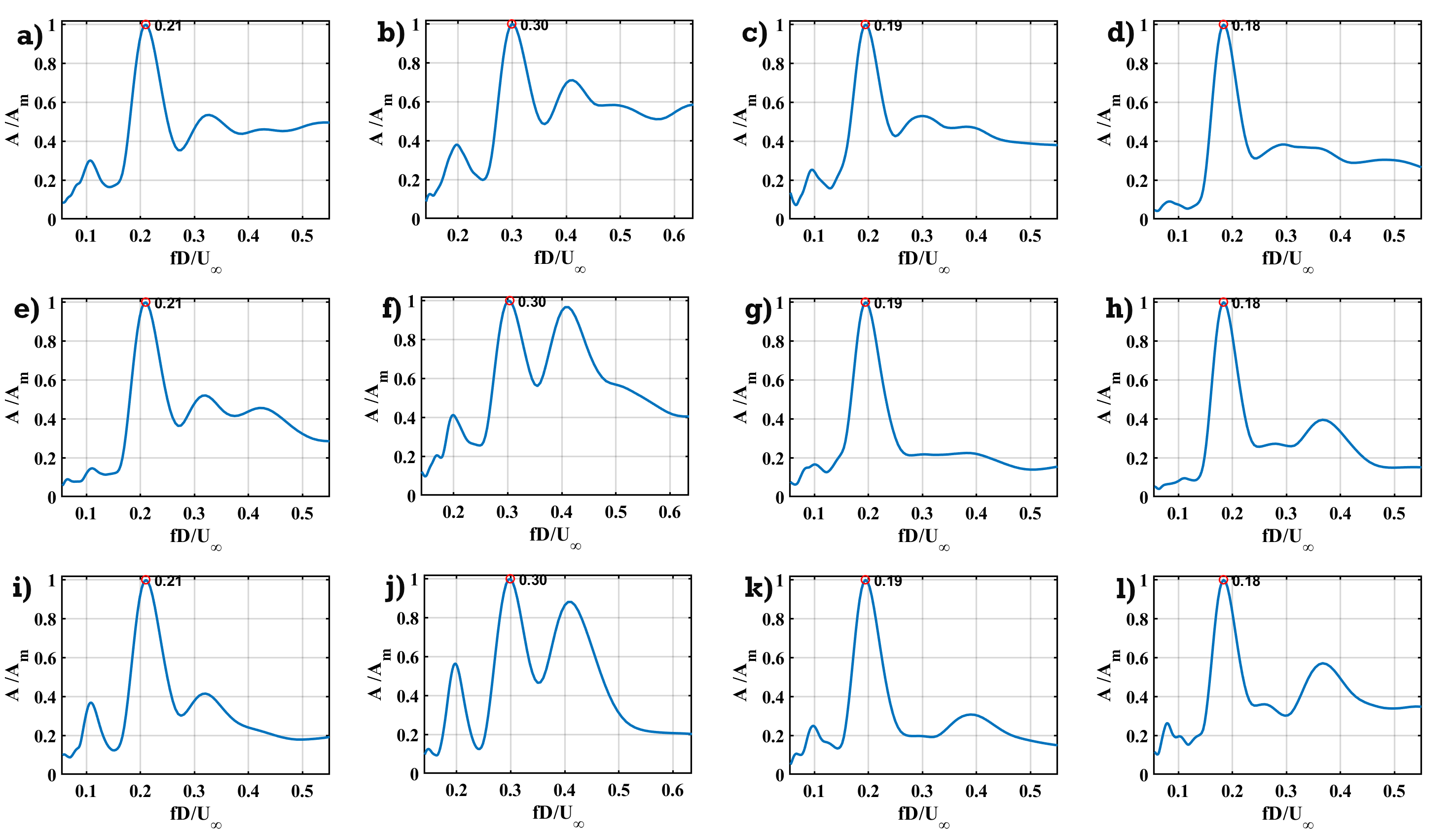} 
   \caption{\label{fig:16} Normalized amplitude (A/ $A_m$)($A_m$ is the maximum amplitude of the system) corresponding to the Strouhal numbers (fD/$U_\infty$) in the probes at the walls of the cavity: (a) without confinement, and with confinement and  (b) no shock impingement, (c) with a flow deflection angle of $8.2^\circ$, and (d) with a flow deflection angle of $13.78^\circ$. For probe rack 1: (e) without confinement, and with confinement and (f) shock impingement, (g) with a flow deflection angle of $8.2^\circ$, and (h) with a flow deflection angle of $13.78^\circ$. For probe rack 2: (i) without confinement, and with confinement and (j) no shock impingement, (k) with a flow deflection angle of $8.2^\circ$, and (l) with a flow deflection angle of $13.78^\circ$.}

\end{figure*}     
The pressure fluctuation at the aft and front walls of the cavity configurations, which take into account the interaction of the shear layer with shocks of different strengths ($44.5^0$ and $ 52.87^0$), are depicted in Figures \ref{fig:10} and \ref{fig:12}. The variation of pressure at the front and aft walls for a complete oscillation cycle is comparable to that of the confined cavity in the absence of shock impingement.

Figure \ref{fig:11} illustrates the spatial variation of density with time for the confined cavity with an impinging shock of strength of $44.5^0$ arising from the top wall deflected at $8.2^0$. The shock that originated at the corner of the deflected surface (3) and is moving in the direction of the shear layer (2) is visible. Before impinging the shear layer, it engages in interaction with the separating shock (4). These two shocks experience a regular reflection, indicated by a purple square in the figures. The upstream traveling wave (1), trailed by the wavefront (5), moves upstream to interact with the separating shear layer (2). During this event, a small vortex is seen rolling up in the floor of the cavity, it also trails the pressure wave and interacts with it before it dissipates after reflection from the front wall (Figures\ref{fig:11}d-i). The upstream traveling pressure wave's interaction with the shear layer, after the former’s reflection from the front wall, is depicted in Figures \ref{fig:11}f–h. In this instance, the shear layer rolls up to impinge on the aft wall while moving downstream and creating vortex. The shock's (3) impingement site on the shear layer is the source of a disturbance (6) (Figure \ref{fig:11}d). The disturbance (6) gets stronger as the perturbations in the shear layer travel downstream (Figure \ref{fig:11}g-i). The shear layer rolls up as a result, and another disturbance (8) is generated which moves in the direction of the upper wall (Figure \ref{fig:11}g-i). As the vortical structures in the shear layer move downstream, they intensify (Figure \ref{fig:11}f-h). Figures \ref{fig:11}h-i show how the disturbance (8) interacts with the separation shock (7) to produce a wave (11), which then interacts with the wavefront of the subsequent cycle (10) that trails the pressure wave (9). The cycle of this configuration, hence,  exhibits the formation of numerous disturbances as a result of the excitement of the impinged shear layer and wave interactions.

The shock from the deflected wall creates some disturbances after it impinges on the shear layer. When the shear layer interacts with the pressure wave inside the cavity after its reflection from the front wall, it is perturbed and some vortices are formed. These vortices become more intense as they pass through the impingement point resulting in the Kelvin- Helmholtz instability. This instability is due to the substantial velocity gradient across the shock, which increases the shear between the consecutive fluid layers. The strength of the shock grows as the deflection angle rises to $13.78^0$. Consequently, as can be observed in the density gradient contours (Figure \ref{fig:13}), the instability increases. Figures \ref{fig:13}a–f illustrate how the upstream traveling pressure wave is reflected as it approaches the front wall. It creates the vortex and disturbs the shear layer. This perturbation in the shear layer further rolls up into more coherent structures as it travels through the impingement point. As they move downstream toward the aft wall, they amplify. Numerous other fluctuations, such as (6) and (7), emanate following the  Kelvin-Helmholtz instability. They interact with the separation shock (8), and the wavefront of the subsequent cycle (10), and eventually combine and migrate to the top wall. Following these interactions, the coalition waves make contact with the boundary layer at the trailing edge.

Figures \ref{fig:9},\ref{fig:11},\ref{fig:13} depict the flow visualization, which indicates that the duration of one complete cycle is the lowest when shock impingement is absent and it increases as the impinging shock strength increases. The rise in the shock strength elevates the Kelvin-Helmholtz instability of the shear layer. Large-scale vortex forms due to this instability, which extracts additional turbulent kinetic energy from the system. Consequently, the energy driving the feedback decreases as a larger part of the energy gets involved in the cascading. This reduces the rapidness of the feedback loop. 
The flow visualizations of all the configurations depict a complete oscillation cycle. There are multiple interactions of shock-shock, shock-boundary layer in most of the configurations, which indicate the presence of numerous sub-events, occurring outside the cavity as well as inside the cavity. We further perform spectral analysis to understand the frequencies of these events.

\subsubsection{\label{sec:level4}Spectral Analysis}
We have installed probes or monitoring stations as described earlier in this article to determine the frequency content of the system. Data collection occurs at a sampling rate of 0.5 GHz, corresponding to a time interval of 2e-09s between each time step. The sampling frequency is sufficiently larger than the frequencies of interest, ensuring compliance with the Nyquist criterion and enabling frequency resolution of a wider range. The number of samples collected for each case is sufficient to provide a frequency resolution of 200 Hz.
\paragraph{\label{sec:level5} Power Spectral Density (PSD)}
 We analyze the power spectral density (PSD) of the unsteady signal to determine the system's frequency content. We implement the 'pwelch' method to obtain the frequency spectrum, applying a Hanning window with a 50 percent overlap. This analysis reveals the frequencies carrying the signal’s energy and quantifies the energy associated with each frequency component. The analysis normalizes the power spectral density (Gxx(p)) using the ratio of the frequency (f) to freestream pressure ($(p_\infty)^2$). It presents the frequency in terms of the non-dimensional Strouhal number (St)(fD/$U_\infty$).

 The normalized PSD from the probes positioned at the middle of the front wall (probe-P1), aft wall (probe-P2), and cavity floor (probe-P3) is shown in the first row of Figure \ref{fig:14}. The second and third rows, which correspond to y = 0.1D (closer to the shear layer) and y = 0.5 D, respectively, were collected from the probes positioned at probe racks 1 and 2. Tables \ref{tab:table1} and \ref {tab:table2} in Section\ref{sec:geo} list the precise locations of the probes in these racks. The variation of the dominant frequencies across configurations and the probe locations is as follows:

\begin{itemize}
\item{In the unconfined cavity (Figures \ref{fig:14} a, e, and i), the dominating Strouhal number is at  \textbf 0.22.The spectrum also shows a higher Strouhal number (St2) at 0.36.}

\item{In the confined cavity (Figure \ref{fig:14} b and f), the dominant Strouhal number is at \textbf{0.32} (St3 = St2 + St1), with the lower Strouhal numbers at 0.11 (St1) and 0.21(St2).}

\item{ The last two columns of Figure \ref{fig:14} depict the spectra for the case of the confined cavity with shock impinging in the separating shear layer at the mid-point. The flow deflection angles are at $8.2^0$ and $13.78^0$ respectively. For the first of these cases (Figure \ref{fig:14} c,g,k), we see the dominant Strouhal number (St1) is at \textbf{0.196}. There is also a mode of higher Strouhal number (St2) of 0.39 but lower power present in the system. For the second case of shock impingement (Figure \ref{fig:14}d,h, and l), we observe the dominant Strouhal number at the second mode (St2) \textbf{0.184} and a lower Strouhal number at 0.08 is visible in the spectra for the probes inside the cavity and higher Strouhal numbers of 0.21(St3) and 0.368(St4) are all also present in the spectra. However, the dominant Strouhal number (St2) has the highest power among all the modes.}

\end{itemize}

The aft wall generates the upstream traveling pressure wave, as shown in the flow fields of the cavities in all configurations (Figures \ref{fig:7},\ref{fig:9},\ref{fig:11},\ref{fig:13} ). As a result, Probe-P2 at the cavity's aft wall has the maximum power spectral density. 
The dominant Strouhal number rises by \textbf{0.1} with the addition of the confinement, as shown in the first two columns of Figure \ref{fig:14}. Figures \ref{fig:7}  and \ref{fig:9}  display the results of the density gradient contours of these two cavity configurations, which show that the feedback loop accelerates when the confinement is present. Similar findings are also shown by the spectral analysis. Additionally, as the shock impingement's strength increases, the dominating frequency drops. Between the confined cavities without shock impingement and those with the impinging shock of maximum strength investigated, there is a noticeable decrease of \textbf{0.136}, which is about $42.5\%$ in the dominating Strouhal number in the latter. The Kelvin-Helmholtz instability causes the energy cascade to be more intense than the feedback loop can support; as a result, the feedback loop slows down due to the lesser energy (Figures \ref{fig:9},\ref{fig:11},\ref{fig:13} ).

These spectra also show that, in comparison to the other probes, probe 6 of probe rack 1, which is situated close to the cavity's trailing edge, has the highest PSD value. In all the cavity arrangements, the aft wall is impacted by the disturbances in the shear layer. Due to its closer proximity to the aft wall, probe 6 on probe rack 1 is exposed to the majority of the energy. The flow visualizations depict the multiple interactions in the mid-region of the shear layer in the domain. These interactions manifest as the highest power in probe 4 at the midpoint of probe rack 2.

\paragraph{\label{sec:level6} Wavelet Analysis}
We have carried out a wavelet analysis to understand the evolution of frequency in terms of Strouhal numbers, with time. Unlike Fourier transformation, wavelet transformation provides great resolution in both frequency and time by allowing us to use scaling and shifting to window data at several resolutions. Wavelets that are stretched (high-frequency content) or compressed (low-frequency content) are superimposed on top of the original signal. Their shapes change with time. The match in the frequency content of the wavelet shape and the original signal establishes a correlation. This process is repeated for each wavelet to identify all the frequencies at every instant. It is recommended that readers consult the works of  Chui \cite{Chui1992}, Torrence \cite{Torrence1998}, and Debnath \cite{Debnath2014} for in-depth explanations and practical uses of wavelet transformation. 
\begin{figure*}[htbp]

	\includegraphics[scale=0.25]{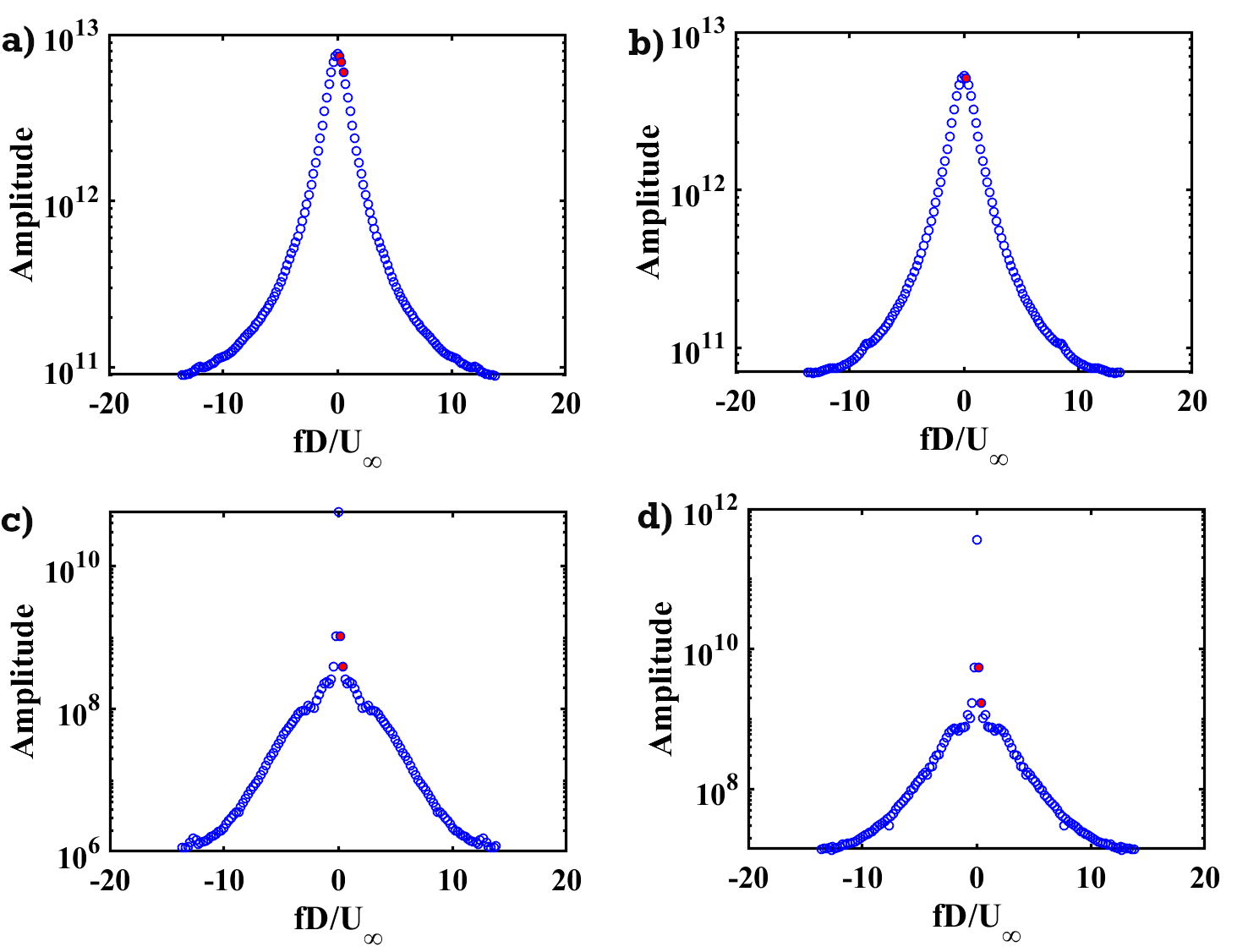} 
    \centering

    \caption{\label{fig:17} Energy associated with the Strouhal numbers for cavity configuration a) without confinement and with confinement b) no deflection angle c) deflection angle at $8.2^0$ d) deflection angle at $13.78^0$.}.

   \end{figure*}
\begin{figure*}[htbp]

	\includegraphics[scale=0.75]{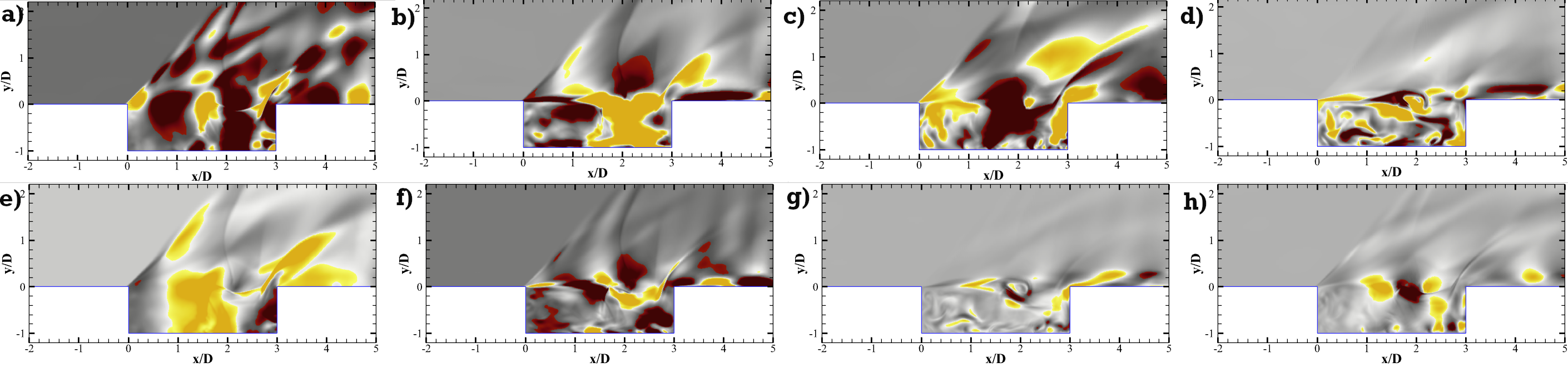} 
  \centering

   \caption{\label{fig:18}  Dynamic Mode 1 computed for a) pressure b) streamwise velocity c) lateral velocity and d) spanwise velocity and Dynamic Mode 2 computed for e) pressure f) streamwise velocity g) lateral velocity and h) spanwise velocity for the unconfined cavity.}.

    \end{figure*}

 \begin{figure*}[htbp]

	\includegraphics[scale=0.75]{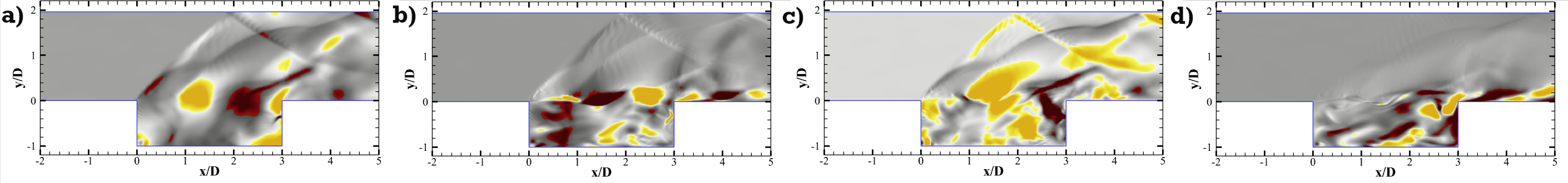}
    \centering

    \caption{\label{fig:19}  Dynamic Mode 1 computed for a) pressure b) streamwise velocity c) lateral velocity and d) spanwise velocity and Dynamic Mode 2 computed for e) pressure f) streamwise velocity g) lateral velocity and h) spanwise velocity for the confined cavity with a flow deflection angle at $0^\circ$ respectively.}.
   \end{figure*}

    \begin{figure*}[htbp]

	\includegraphics[scale=0.75]{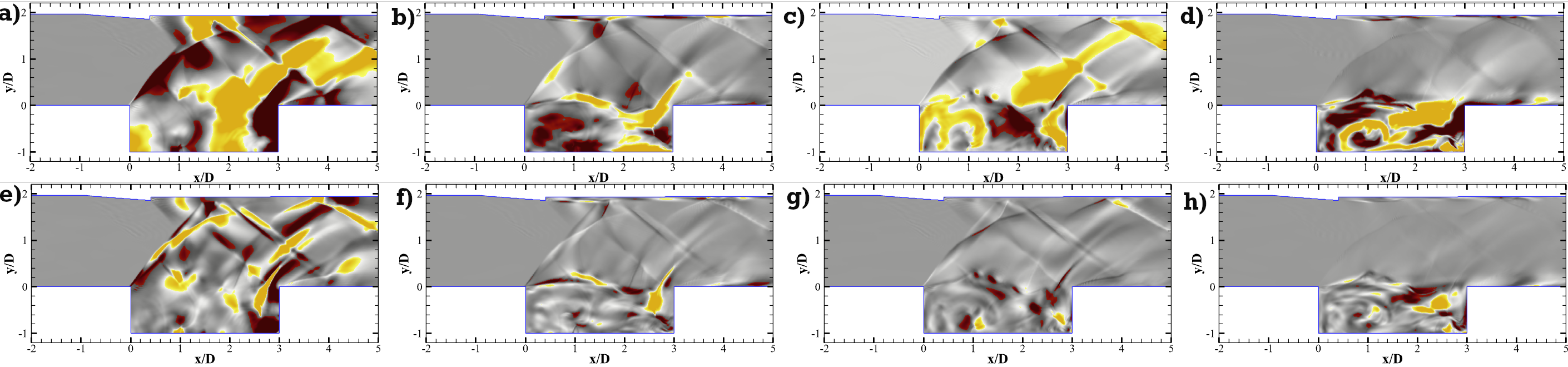}
    \centering

    \caption{\label{fig:20} Dynamic Mode 1 computed for a) pressure b) streamwise velocity c) lateral velocity and d) spanwise velocity and Dynamic Mode 2 computed for e) pressure f) streamwise velocity g) lateral velocity and h) spanwise velocity for the confined cavity with a flow deflection angle at $8.2^\circ$ respectively.}.

   \end{figure*}

    \begin{figure*}[htbp]

	\includegraphics[scale=0.75]{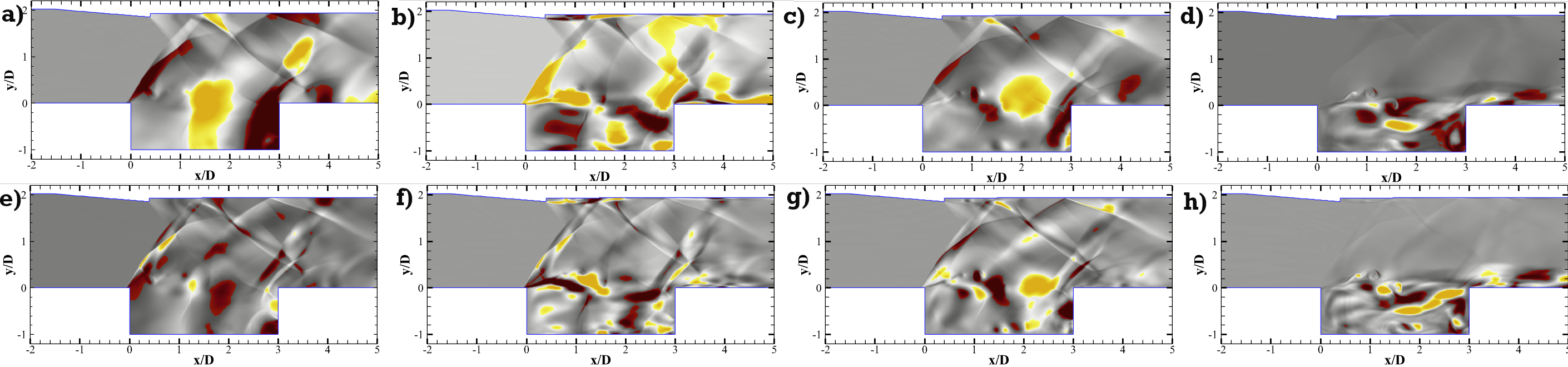}
    \centering

    \caption{\label{fig:21} Dynamic Mode 1 computed for a) pressure b) streamwise velocity c) lateral velocity and d) spanwise velocity and Dynamic Mode 2 computed for e) pressure f) streamwise velocity g) lateral velocity and h) spanwise velocity for the confined cavity with a flow deflection angle at $13.78^\circ$ respectively.}.

   \end{figure*}

The turbulent flow field produces a non-stationary pressure signal with transient features. Unlike the spectrogram's fixed window size, CWT uses wavelets at varying scales, offering a better time and frequency resolution. Hence, it is more effective for capturing transient, non-stationary, and oscillatory events in the flow. Consequently, in the current investigation, we have analyzed a signal's temporal variation of frequency content using the Continuous Wavelet Transform (CWT). The process entails convolving the signal through a family of wavelets produced through the translation and scale of a mother wavelet. Furthermore, we conducted a Continuous Wavelet Transform (CWT) analysis with a temporal average. Even with PSD and CWT data, this analysis is essential because it provides information on the dominant frequencies that are continuously present throughout the signal length, complementing the global frequency content provided by the PSD and the instantaneous information provided by the CWT. 

The wavelet transformation of the signal from the probes in all configurations with the highest normalized PSD content is shown in Figure \ref{fig:15}. For all the arrangements, these spectra yield a dominant Strouhal number, which is comparable to the one derived from the power spectral density. Furthermore, high-energy zones (shown by the red blotches in all the figures) that correspond to higher Strouhal numbers are visible. In contrast to PSD analysis, wavelet analysis allows us to capture all frequency modes with high frequency and temporal resolutions. These higher energy zones indicate transient events that occurred more rapidly than those associated with the dominant frequency. Several shock-shock and shock shear interactions, which are some transitory occurrences, are seen in the flow visualizations. The wavelet studies have identified these interactions as high-energy events. Notably, the last two columns of Figure \ref{fig:15} corresponding to the confinement with deflection angles, show how these occurrences amplify in the presence of shock-shear interaction. This is evident in the continuous high-energy blotches at specific frequencies throughout the period.

The time-averaged wavelet transformation of the signal collected from probes 6 and 4 in probe racks 1 and 2, respectively, is shown in Figure \ref{fig:16}. It shows the amplitude (A) normalized with the maximum amplitude ($A_m$) corresponding to the Strouhal numbers in the system. Similar to the probe near the cavity's aft wall, these probes are selected for additional analysis because they contain the greatest normalized PSD values. We obtain dominant Strouhal numbers that resemble those in the PSD plots.

The signal's Fourier transformation yields reliable information on the system's frequency content, indicated by the Strouhal numbers. Comprehensive frequency analysis across scales and positions is made possible by wavelet transformation, and this is essential for identifying transient events with short durations that are defined by localized high-energy zones. These spectrum studies, in conjunction with the flow visualizations of the cavity configurations, allow us insight into the frequency content of the system, its temporal evolution, and the events that take place within the cavity flow field during that period. We will conduct a more detailed dynamic mode decomposition analysis to gain a deeper insight into the relationship between the system's frequencies and the flow features.



 \begin{figure*}

	\includegraphics[scale=0.9]{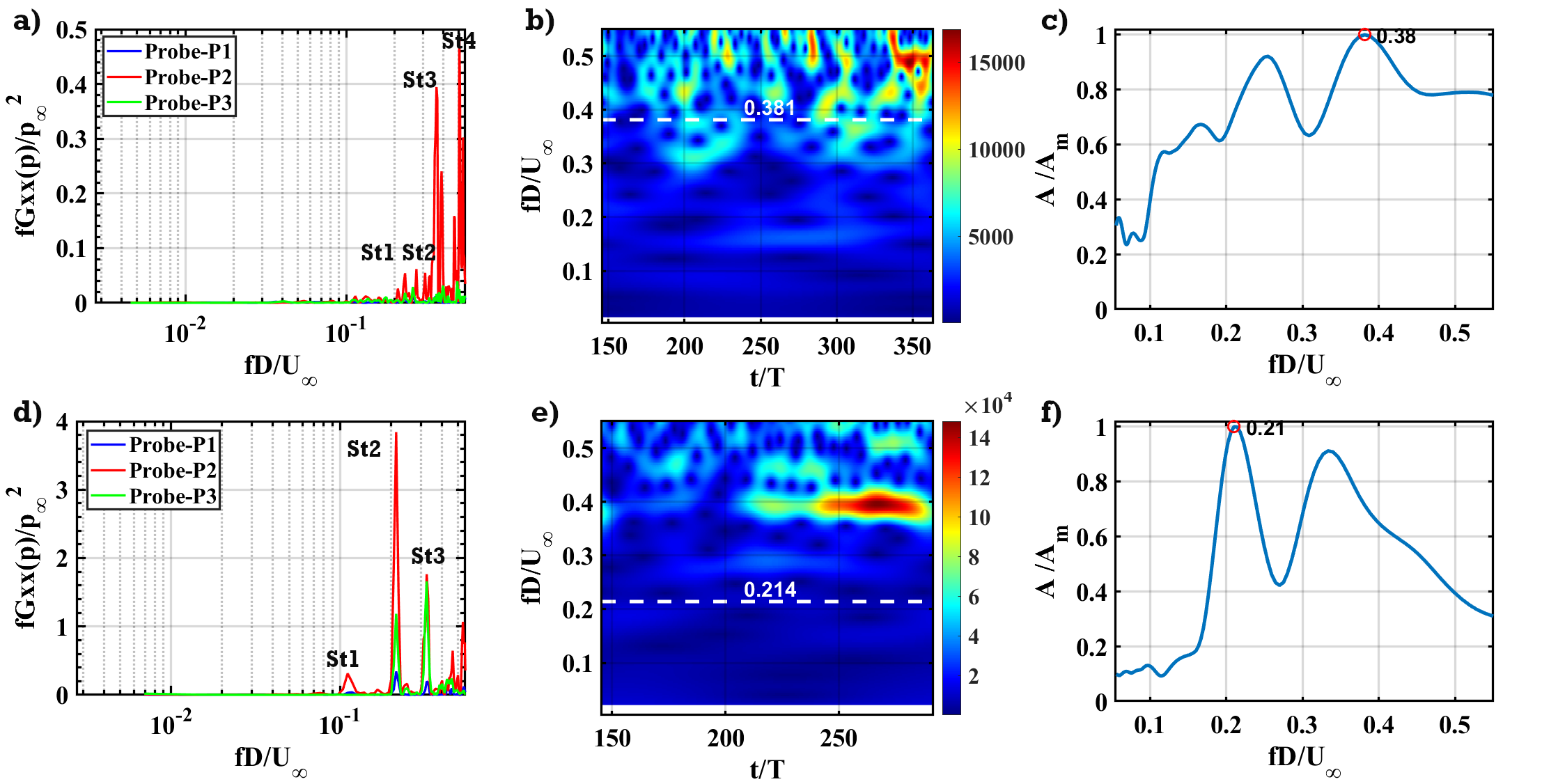}
    \centering

    \caption{\label{fig:22} Normalized Power Spectral Density (fGxx(p)/$(p_\infty)^2$ vs (fD/$U_\infty$), Continuous Wavelet transformation (fD/$U_\infty$) and Normalized amplitude (A/$A_m$)corresponding to Strouhal numbers(fD/$U_\infty$) for the confined cavity a-c) without and d-f) with shock-shear interaction respectively, at Mach number 2. }

   \end{figure*}
\begin{figure*}

	\includegraphics[scale=0.31]{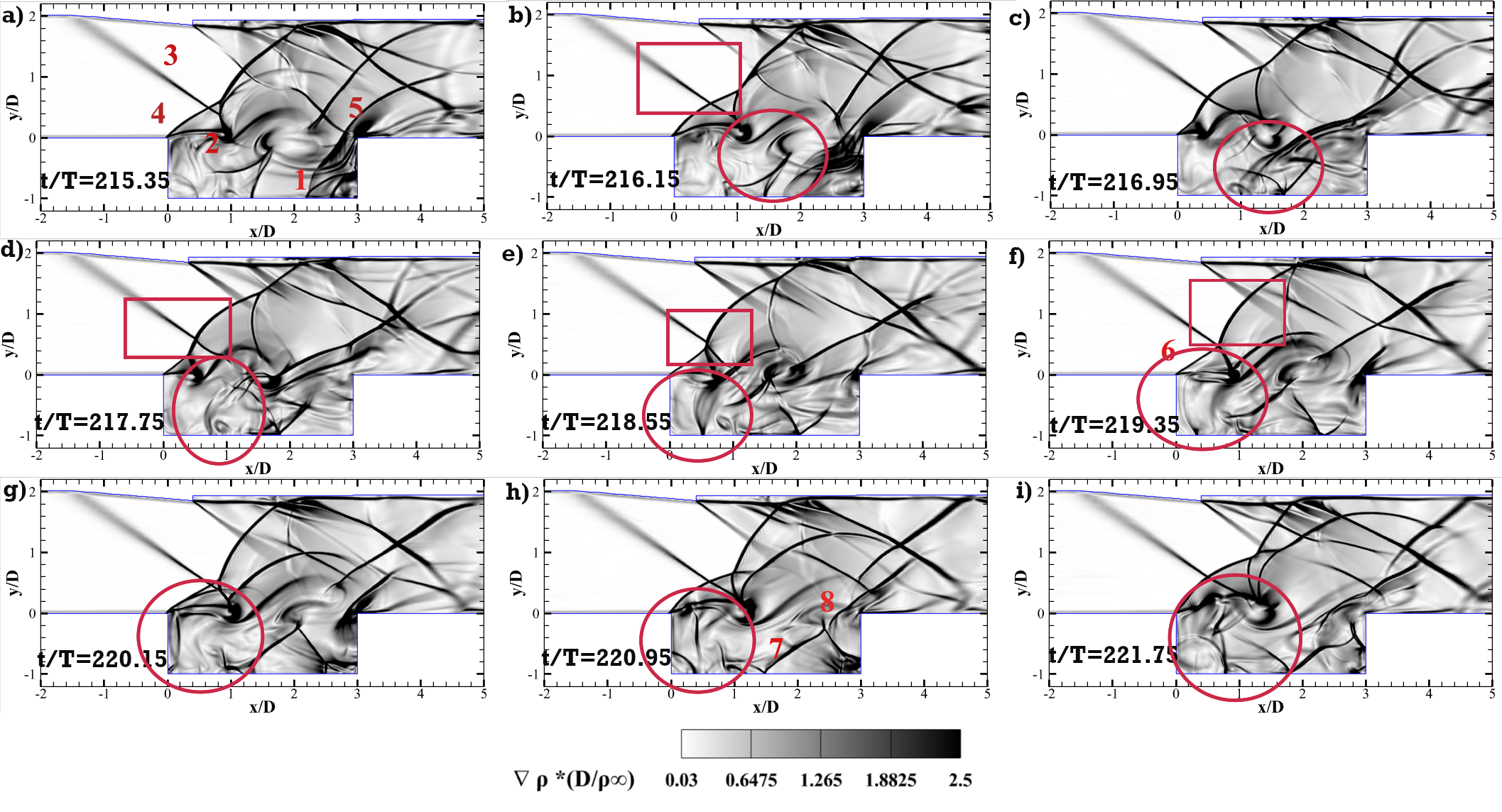} 
    \centering

    \caption{\label{fig:23} Normalised Density Gradient  ($\nabla\rho*(D/\rho_\infty$)) of the cavity with the top wall deflected at $13.78^0$  for one complete cycle from the time step (t/T) 215.35 (a) to 221.75 (i) at an interval of 0.8. The salient features in the flow field are marked as: (1) the upstream traveling pressure waves (2)the K-H instability rolls in the perturbed shear layer, (3) the impinging shock wave from the deflected top wall, (4) the separation shock at the leading edge, (5) trailing wavefront, (6) separation shock of the next cycle, (7) \& (8) the upstream traveling wave and the external wavefront trailing it for the next cycle.  The red circles show the features inside the cavity. The red illustrates the multiple events outside the cavity.}
   \end{figure*}

\subsubsection{\label{sec:level7}Dynamic Mode Decomposition}
Dynamic Mode Decomposition (DMD) is a data-driven technique for extracting spatio-temporal coherent structures from time-resolved data to analyze the dynamics of linear and non-linear systems \cite{rowley2009,schmid2010,tu2014,jovanovic2014,kutz2016,brunton2016}.  A thorough yet simplified depiction of the dynamics of the system is provided by DMD, which breaks down the complex system into dynamic modes with distinct growth rates and frequencies represented by the eigenvalues and the eigenvectors. These eigenvalues and eigenvectors are useful in fluid dynamics analysis and identification of coherent structures including wave packets, shear layers, and vortices.

In our investigation, we carry out the DMD analysis by the parallel QR-decomposition and Singular Value Decomposition (SVD) to obtain the dynamic modes \cite{sayadi2016parallel,soni2019modal,arya2021effect,bhatia2019numerical,sharma2024investigation,arya2023acoustic}.The number of the snapshots, that we use in the study satisfy the Nyquist criteria and achieve the necessary frequency resolution. For the cases under study, with a frequency resolution of 200 Hz and the maximum frequency of 20kHz, that we try to resolve , the number of the snapshots is 200. This approach ensures that our analysis captures all significant frequencies.   After arranging snapshot data into matrices, we execute a QR decomposition  to obtain an orthogonal matrix (O) and an upper triangular matrix(R). Further SVD decomposes the upper-triangular matrix, 
R, obtained from QR-factorization and perform a linear mapping between the two corresponding snapshots to obtain the eigenvalues and vectors. These two decompositions reduce the dimensionality of the system and subsequently provide a computationally economic foundation for further analysis of the dynamics \cite{sayadi2016parallel}. The eigenvalues and eigenvectors of the resulting reduced-order matrix contain information about the frequencies and the corresponding flow features of the system. The spatial structures that each eigenvalue is associated with, are represented by the dynamic modes, which offer insights into the coherent structures existing in the data as they change. 


Figure \ref{fig:17} presents the spectra for the DMD, with Strouhal numbers considered till a maximum frequency limit of 20 kHz.  The Strouhal numbers corresponding to these frequencies have a higher amplitude than the higher ones. Consequently, these modes represent the significant flow features and hence, the pressure and velocity fields corresponding to these modes are studied further.

In the case of the unconfined cavity, Figure \ref{fig:18} displays the dominant dynamic mode,as seen in the DMD spectra (Figure \ref{fig:17}a), corresponding to Strouhal number 0.22 and the higher Strouhal number at 0.36 derived for the pressure field and the three components of velocity in the two corresponding rows. The spectral analysis of this cavity reveals a prominent Strouhal number of 0.22. The wavelet analysis (Figures \ref{fig:15}a, e, and i) yields higher energy at some higher Strouhal number of 0.4, which indicates the second dynamic mode. Large coherent structures are visible in both the cavity and the shear layer from the first dynamic mode (Figures \ref{fig:18} a-d). The shear layer separates from the leading edge and interacts with the traveling wave upstream, forming coherent patterns. These structures in the shear layer travel downstream and impinge on the trailing edge. Moreover, the mass entering the cavity creates vortices that cause an upstream traveling wave to form close to the aft wall. These are the principal characteristics of a cavity's feedback loop. The first mode represents the large-scale structures inside the cavity and the shear layer, which are a part of the feedback loop. The higher mode captures the minor structures found in the shear layer and the cavity (Figure \ref{fig:18} e–h) as a result of the different fluctuations brought about by the shear layer's perturbation and the interaction of the separation shocks from the subsequent cycles.

Figure \ref{fig:19} displays the dynamic mode derived for the pressure and velocity fields for the confined cavity, with no shock-shear layer interaction. Probes at the cavity walls and probe rack 1 close to the shear layer have a dominant mode at 0.32 in the spectral analysis (Figure \ref{fig:14},\ref{fig:15},\ref{fig:16}). The DMD spectra also show this Strouhal number has a higher amplitude (Figure \ref{fig:17} b). This dynamic mode is associated with vortices in the shear that are convecting downstream (Figure \ref{fig:19} b), upstream traveling pressure wave within the cavity (Figure \ref{fig:19} c), the separating shock as well as its reflection from the top wall (Figures \ref{fig:19} a-c). 

 Figure \ref{fig:20} depicts the two dynamic modes computed for pressure and the three components of the velocity in the two corresponding rows in the case of the confined cavity with a deflection angle of $8.2^0$. The contours in the first row are of the pressure field while the streamwise, lateral, and spanwise velocities are represented by the next three-row contours. As seen from the corresponding DMD spectra Figure \ref{fig:17}c, the first mode corresponds to a Strouhal number of 0.196, obtained in the PSD, wavelet, and the time-averaged wavelet analyses as the dominant. The second mode is associated with a Strouhal number of 0.39, which is seen as the high energy blotches in the wavelet analysis (Figures \ref{fig:15} c,g, and k). We can observe from Figures \ref{fig:20}a–d that the large-scale coherent structures found in the system are associated with the dominant frequency mode as determined by the spectral analysis. The large-scale structures found in the disturbed shear layer (figure \ref{fig:20}b), the upstream traveling wave's impingement and interaction with the shear layer (figure \ref{fig:20}c), and the vortex roll-up in the cavity's base that follows the upstream traveling wave (figure \ref{fig:20}d) are all depicted in the velocity contours corresponding to this mode. Figure \ref{fig:20}c also shows the generation of the vortices that initiate the traveling wave upstream at the aft wall. The separation shock at the leading edge is visible in the contours corresponding to this mode. Figures \ref{fig:20}e–h representing the higher frequency mode, depict how the shocks interact with various external disturbances outside the cavity. The small structures present in the rolling up of the shear layer after interaction with the upstream traveling wave after its reflection from the leading edge are also prominent (Figure \ref{fig:20}g). As a result, the feedback loop is linked to the dominant frequency mode, or lower dynamic mode, whereas the other disturbances and their interactions with the top wall and separation shocks are associated with the higher dynamic mode.

 Figure \ref{fig:17}d shows two dynamic modes at Strouhal numbers 0.184 and 0.36 having the higher amplitude. The spectral analysis illustrates the Strouhal number 0.184 is the dominating mode with the higher number of 0.36, corresponding to higher energy structures existing for a smaller duration in the wavelet analysis.Figure \ref{fig:21} shows the three components of the velocity in the two corresponding rows (Figures \ref{fig:21}b-d and f h) as well as the two dynamic modes computed for the pressure field (Figures \ref{fig:21}a and e) in the configuration of the confined cavity with a deflection angle of $13.78^0$.  Likewise the previous case, the dominant frequency is associated with large coherent structures related to the bulging out of the shear layer in interaction with the upstream traveling waves (Figure \ref{fig:21} b), and the large-scale vortices in the excited shear as it travels downstream (Figure \ref{fig:21} c-d). These structures signify the feedback loop in the system. The higher frequency mode mainly corresponds to the small structures in the shear layer due to the perturbation (Figure \ref{fig:21}f-g), and the multiple interactions in the domain arising from the disturbances, the separation and reflected compression waves.

We observe the modification of the flow characteristics of the cavity in a cavity with and without the top-wall confinement and the impact of shock impingement in the shear layer from this section \ref{effect1}. From all the analyses, it is seen that the confined cavity with no shock impingement has the highest frequency whereas the confined cavity with impingement of the shear layer with shock of the highest strength, has the least frequency. Therefore, we will examine the effect of Mach number in this frequency reduction in the next section. We initiate the flow with a larger velocity to increase the flow Mach number to 2.

\subsection{\label{effect2}Effect of Mach number on the confined cavity characteristics in the presence and absence of shock-shear layer interaction .}
\subsubsection{\label{sec:spectra2} Spectral Analysis}
The spectra analysis of the confined cavity with and without shock impingement for the three probes positioned in the middle of the cavity walls at Mach number 2, is displayed in Figure \ref{fig:22}. The PSD, wavelet, and time-averaged wavelet analysis show that, in the absence of shock impingement, the dominating Strouhal number (St4) is \textbf{0.381}(Figure \ref{fig:22}a-c). The dominating Strouhal number (St2) of the confined cavity with the impinging shock of the deflection angle of $13.78^0$ is \textbf{0.21} (Figure \ref{fig:22}d-f). The dominant frequencies for both configurations have increased with the Mach number as the overall speed of the flow has increased, which speeds up the events. However, this increment is more pronounced in the absence of shock-shear layer interaction. The reduction in Strouhal number in the cavity configuration with shock-shear layer interaction increases to almost \textbf{0.167}, with the increased Mach number. This reduction is \textbf{22.7$\%$}  more than in the case of Mach number of 1.71 (Figures \ref{fig:14}b and d). The genesis of the Kelvin-Helmholtz instability is connected to the greater decrease in frequencies caused by shock impingement as seen in Section \ref{effect1}. We shall examine the flow visualization of the confined cavity with the shock impingement to comprehend the aforementioned observations.

\subsubsection{\label{sec:flow2} Flow Field Visualisation}
Figure \ref{fig:23}shows that when the Mach number increases, there are more disruptions in the flow field of this configuration. The separating shear layer(2) comprises coherent structures, which intensify while moving downstream. Due to the bulging of the shear layer, the point where the shock (3) impinges also oscillates(Figure \ref{fig:23}a-c). Before interacting with the shear layer, the separation shock (4) and the impinging shock (3) interact at the leading edge. Figures \ref{fig:23} a-i show the formation, upstream motion, and reflection of the pressure wave (1) inside the cavity with its trailing wavefront (5). This wave front interacts with the coherent structures of the shear layer as the upstream traveling wave meets the downstream traveling shear layer excited from the previous cycle. This generates disturbances that merge with the separation shock (6) forming as the incoming shear layer again detaches from the leading edge (Figures \ref{fig:23}e-f). Coherent structures emerge as the shear layer interacts with the pressure wave after it reflects from the front wall (Figures \ref{fig:23}g-f). Kelvin-Helmholtz instability forms in the shear layer as the vortices roll up. The instability intensifies as the perturbation pass through the impingement point (Figure \ref{fig:23}g-h). Further, as the coherent structures in the shear layer travel towards the aft wall, they encounter the pressure wave of the next cycle (7), which is traveling upstream (Figure \ref{fig:23}i). This flow field shows that the instability is more prominent than what we observed in the Mach 1.71 scenario (Figure \ref{fig:13}). More energy cascades out of the system as a result of the extreme instability amplifying the turbulence. As a result, less energy is used to drive the cavity's feedback loop, which lowers the increment in the oscillation frequency with an increase in Mach number.

 \section{\label{sec:conclusion}Conclusion}
The current study examines a supersonic flow past an open cavity with an L/D ratio of 3, both in the presence and absence of top wall confinement. The effect of the shock of varying strength impinging the shear layer separating from the leading edge, on the flow field of the confined cavity is investigated. Lastly, the effect of Mach number along with the shock impingement in the confined cavity is analyzed. The important conclusions from our study are:
 \begin {itemize}
 \item{Flow visualizations deduce that the period for one complete cycle inside the cavity is less in the case of the top wall confinement. Spectral investigations reveal that the cavity's confinement raises the system's dominant frequency. The pressure's temporal change indicates that there is a greater reduction in pressure at the aft wall when the top wall confinement is present than in its absence.  This enhances the mass inflow through the aft wall of the cavity and accelerates the feedback loop, which further raises the dominant frequency.}
 \item{For a Mach number of 1.71, the spectral analysis additionally reveals that the dominant Strouhal number is declining by \textbf{42.5\%} as the impinging shock strength rises. This reduction is attributed to the Kelvin-Helmholtz instability, which causes perturbations in the shear layer to roll up as they pass through the impingement point after interacting with the reflected wave from the leading edge. The instability amplifies the turbulent kinetic energy, thereby reducing the energy available to sustain the feedback loop within the cavity. This phenomenon is effectively captured by the dynamic modes and the spatio-temporal variations in density. }
 
 \item{In the confined cavity, an increase in Mach number has raised the overall dominant frequency; however, in the event of shock impingement, this increment is less pronounced. The increase in Mach number to 2, increases the Strouhal number by \textbf{16$\%$} in the presence of shock impingement while it is near \textbf {19.06$\%$} in the absence of shock impingement. The decrease in Strouhal number in the presence of the shock impingement at this increased Mach number is \textbf{0.167} than in the absence of it, which is \textbf{22.7$\%$}   more than that obtained at Mach number 1.71.  The density gradient contours indicate that the instability in the shear layer intensifies as the Mach number increases. Consequently, turbulence extracts more energy from the system, diminishing the energy available to drive the feedback loop. This results in a smaller increase in oscillation frequency with the rising Mach number in the presence of shock-shear layer interaction.}
 \end{itemize}
   
\clearpage

\begin{acknowledgments}
The authors acknowledge the National Supercomputing Mission (NSM) for providing computing resources of 'PARAM Sanganak' at IIT Kanpur, which is implemented by C-DAC and supported by the Ministry of Electronics and Information Technology (MeitY) and Department of Science and Technology (DST), Government of India. Also, we would like to thank the computer center (www.iitk.ac.in/cc) at IIT Kanpur for providing the resources to carry out this work.
\end{acknowledgments}

\section*{Conflict of Interest}
The authors have no conflict of interest to disclose.

\section*{Data Availability}
The data that support the findings of this study are available from the corresponding author upon reasonable request.


\section{Appendixes}
\subsection{Nomenclature}
\begin{table}[htbp] \caption{Symbols and their corresponding descriptions.}
    \centering
    \begin{ruledtabular}
    \begin{tabular}{ll}
        \textbf{Symbol} & \textbf{Description} \\
        \hline
        D   & Depth of the cavity \\
        L   & Length of the cavity \\
        $\beta$ & Shock angle \\
        $\theta$ & Deflection angle \\
        $H_i$ & Height of the inlet \\
        $H_o$ & Height of the outlet \\
        $p_\infty$ & Freestream pressure \\
        $U_\infty$ & Freestream velocity \\
        $<p>$ & Average pressure \\
        s & Radial distance \\
        $\rho_\infty$ & Freestream density \\
        $\nabla\rho$ & Gradient of density  \\
        t & Time \\
        T & Reference time \\
        f & Frequency \\
        $\frac{fD}{U_\infty}$ & Strouhal number(St) \\
        A & Amplitude \\
        $A_m$ & Maximum amplitude of the system \\
        Gxx(p) & Power Spectral Density \\
    \end{tabular}
    \end{ruledtabular}
   
    \label{tab:symbols}
\end{table}







\section*{References}
\nocite{*}
\bibliography{aipsamp}

\end{document}